\documentclass[12pt]{article}
\usepackage{a4wide}
\usepackage{latexsym}
\usepackage{amsmath}
\usepackage{amssymb}
\usepackage{amsfonts}
\usepackage{amscd}
\usepackage{shuffle}
\usepackage{cite}
\usepackage{graphicx}
\usepackage{float}
\usepackage{axodraw}

\usepackage{pslatex}
\usepackage[latin1,utf8]{inputenc}
\usepackage[OT2,T1]{fontenc}

\usepackage{ifthen}

\allowdisplaybreaks

\newcommand{\bq}{\begin{eqnarray}}
\newcommand{\eq}{\end{eqnarray}}
\newcommand{\eps}{\varepsilon}

\begin{document}

\thispagestyle{empty}

\begin{flushright}
  Edinburgh 2017/26 \\
  MITP/17-091
\end{flushright}

\vspace{1.5cm}

\begin{center}
  {\Large\bf Properties of scattering forms and their relation to associahedra\\
  }
  \vspace{1cm}
  {\large Leonardo de la Cruz ${}^{a}$, Alexander Kniss ${}^{b}$ and Stefan Weinzierl ${}^{b}$\\
  \vspace{1cm}
      {\small ${}^{a}$ \em Higgs Centre for Theoretical Physics, School of Physics and Astronomy,}\\
      {\small \em The University of Edinburgh,}\\
      {\small \em Edinburgh EH9 3JZ, Scotland, UK}\\
\vspace{2mm}
      {\small ${}^{b}$ \em PRISMA Cluster of Excellence, Institut f{\"u}r Physik, }\\
      {\small \em Johannes Gutenberg-Universit{\"a}t Mainz,}\\
      {\small \em D - 55099 Mainz, Germany}\\
  } 
\end{center}

\vspace{2cm}

\begin{abstract}\noindent
  {
We show that the half-integrands in the CHY representation of tree amplitudes give rise to the definition of 
differential forms -- the scattering forms --
on the moduli space of a Riemann sphere with $n$ marked points.
These differential forms have some remarkable properties.
We show that all singularities are on the divisor
$\overline{\mathcal M}_{0,n} \backslash {\mathcal M}_{0,n}$.
Each singularity is logarithmic and the residue factorises into two differential forms of lower points.
In order for this to work, we provide a threefold generalisation of the CHY polarisation factor 
(also known as reduced Pfaffian)
towards off-shell momenta, unphysical polarisations and away from the solutions of the scattering equations.
We discuss explicitly the cases of bi-adjoint scalar amplitudes, Yang-Mills amplitudes and gravity amplitudes.
   }
\end{abstract}

\vspace*{\fill}

\newpage

\section{Introduction}
\label{sect:intro}

In this article we bring three things together, which really should be viewed together:
(i) the Cachazo-He-Yuan (CHY) representation of tree-level $n$-point scattering amplitudes \cite{Cachazo:2013gna,Cachazo:2013hca,Cachazo:2013iea}, 
(ii) the moduli space ${\mathcal M}_{0,n}$
of $n$ marked points on a Riemann surface of genus zero and
(iii) ``positive'' geometries / ``canonical'' forms, 
as recently discussed by Arkani-Hamed, Bai and Lam \cite{Arkani-Hamed:2017tmz}.
The integrand of the CHY representation for bi-adjoint scalar amplitudes, Yang-Mills amplitudes and gravity amplitudes 
is constructed from two factors, a cyclic factor (or Parke-Taylor factor) and a polarisation factor (also known as reduced Pfaffian).
We show that the cyclic factor and the polarisation factor lead to 
differential $(n-3)$-forms $\Omega_{\mathrm{scattering}}^{\mathrm{cyclic}}$ and $\Omega_{\mathrm{scattering}}^{\mathrm{pol}}$, respectively, 
on the compactification $\overline{\mathcal M}_{0,n}$
of ${\mathcal M}_{0,n}$, such that the only singularities of
the differential forms $\Omega_{\mathrm{scattering}}$
are on the divisor $\overline{\mathcal M}_{0,n} \backslash {\mathcal M}_{0,n}$.
Each singularity is logarithmic and the residue factorises into two differential forms of lower points.
These scattering forms figure prominently in the recent work by Mizera \cite{Mizera:2017rqa,Mizera:2017cqs}.
The scattering forms are cocycles and Mizera has shown that the amplitudes are intersection numbers
of these cocycles, twisted by a one-form derived from the scattering equations.

We put ``positive'' geometry into quotes.
The reason is the following: The solutions of the scattering equations are in general complex
and correspond to points in ${\mathcal M}_{0,n}$.
Only for very special external momenta $p$ are the solutions of the scattering equations real \cite{Cachazo:2016ror,Weinzierl:2014vwa,Kalousios:2013eca}.
If the solutions are real, we may limit ourselves to the space of real points ${\mathcal M}_{0,n}({\mathbb R})$.
This is a positive space in the sense of Arkani-Hamed, Bai and Lam \cite{Arkani-Hamed:2017tmz}, with boundary 
$\overline{\mathcal M}_{0,n}({\mathbb R}) \backslash {\mathcal M}_{0,n}({\mathbb R})$.
However, we are interested in the general situation.
This forces us to work throughout the paper with the complex numbers ${\mathbb C}$ instead of the real numbers ${\mathbb R}$.
For simplicity we write ${\mathcal M}_{0,n}$ instead of ${\mathcal M}_{0,n}({\mathbb C})$.
We find that the notion of ``positivity'' is not essential, what is essential is the structure of 
the divisor $\overline{\mathcal M}_{0,n} \backslash {\mathcal M}_{0,n}$, which generalises in a straightforward way
from the real case $\overline{\mathcal M}_{0,n}({\mathbb R}) \backslash {\mathcal M}_{0,n}({\mathbb R})$
towards the complex case $\overline{\mathcal M}_{0,n}({\mathbb C}) \backslash {\mathcal M}_{0,n}({\mathbb C})$.

We also put ``canonical'' form into quotes.
Here, our reason is as follows:
The word ``canonical'' implies, that the differential form is unique for a given geometry.
For example, following \cite{Arkani-Hamed:2017tmz} the differential form for $3$ external particles is a $0$-form 
and should for positive geometries be equal to $\pm 1$, depending on the orientation.
For pseudo-positive geometries the value zero is also allowed.
We find this to be the case for $\Omega_{\mathrm{scattering}}^{\mathrm{cyclic}}$.
However, we also have $\Omega_{\mathrm{scattering}}^{\mathrm{pol}}$, which in the three-point case reduces to the
$0$-form given by the three-point amplitude (up to a factor $\pm i$,
depending on the orientation).
As the geometry of the space always stays the same (we always look at ${\mathcal M}_{0,n}$),
we are forced to give up the requirement that for $n=3$ the differential $0$-form takes the values $\{-1,0,1\}$ and allow
the more general situation that for $n=3$ the differential $0$-form is given by the corresponding three-point amplitude 
(up to a factor $\pm i$, depending on the orientation).

There is a second -- and a slightly more subtle -- reason why we put ``canonical'' into quotes.
As mentioned above, for three external gauge bosons the $0$-form $\Omega_{\mathrm{scattering}}^{\mathrm{pol}}$
is given (up to simple proportionality constants) by the three-point amplitude.
This is supplemented by an infinite (countable) tower 
of additional $0$-forms involving auxiliary particles, such
that in the factorisation at $n$-points only a finite number of additional $0$-forms occur.
The additional $0$-forms vanish for on-shell kinematics, but are non-zero for off-shell kinematics.
They are closely related to the construction of BCJ-numerators from an effective Lagrangian.
It is known that starting from $n=5$ such an effective Lagrangian is not unique.
This implies that for $n \ge 5$ there is more than one possibility 
to define $\Omega_{\mathrm{scattering}}^{\mathrm{pol}}$.

The CHY representation of Yang-Mills amplitudes and gravity amplitudes involves one or two polarisation factors $E(p,\eps,z)$.
The standard definition of the polarisation factor $E(p,\eps,z)$ (for on-shell momenta $p$, transverse polarisations $\eps$ and values $z$
satisfying the scattering equations)
is given in terms of a reduced Pfaffian.
We will need a threefold generalisation of the polarisation factor:
towards off-shell momenta $p$, unphysical polarisations $\eps$ and general values $z \in {\mathcal M}_{0,n}$, not restricted to the zero-dimensional
sub-variety defined by the scattering equations.
We give a definition of the polarisation factor for this general case,
which agrees with the reduced Pfaffian for on-shell momenta, transverse polarisations and on the sub-variety defined by the scattering equations.

Let us mention that shortly after our paper appeared on the arXive, ref.~\cite{Arkani-Hamed:2017mur} appeared on the arXive, 
where scattering forms on the kinematic space of Mandelstam variables are studied.
Ref.~\cite{Arkani-Hamed:2017mur} defines a kinematic associahedron, where positivity conditions 
for all planar Lorentz invariants $s_{i_1,...,i_k}$
and constraints $-s_{ij}=c_{ij}>0$ for non-adjacent Lorentz invariants with $1 \le i < j \le (n-1)$ are imposed.
With these constraints, the scattering forms on the kinematic space are related by a push-forward
to the scattering forms discussed in our paper.
In this paper we do not impose the positivity conditions nor the constraints $-s_{ij}=c_{ij}>0$.

This paper is organised as follows:
In section~\ref{sect:review} we introduce our notation and review a few basic concepts.
The moduli space ${\mathcal M}_{0,n}$ of $n$ marked points on a Riemann surface of genus zero
and its compactification $\overline{\mathcal M}_{0,n}$
plays a prominent role in this paper and we review the definition and essential properties in section~\ref{sect:M_0_n}.
Section~\ref{sect:scattering_forms} contains the main part of this paper. We study
the scattering forms and exhibit some remarkable properties.
Our conclusions are contained in section~\ref{sect:conclusions}.
In the appendix we collected useful information on the standard definition of the reduced Pfaffian and technical details on proofs.

\section{Review of basic facts}
\label{sect:review}

\subsection{Notation}

Let us consider a scattering process with $n$ massless particles in $D$ space-time dimensions
within the Born approximation.
We denote the momenta of the external particles by $p=\{p_1,...,p_n\}$.
By convention we take all momenta to be outgoing, momentum conservation reads therefore
\bq
 \sum\limits_{i=1}^n p_i & = & 0.
\eq
If all external particles are gauge bosons, we denote by
$\eps=(\eps_1,...,\eps_n)$ the polarisation vectors.

Let $I$ be a subset of $\{1,2,...,n\}$.
We set
\bq
 p_I 
 & = & 
 \sum\limits_{i \in I}
 p_i
\eq
and
\bq
 s_I
 & = &
 p_I^2.
\eq
Let $\sigma=(\sigma_1,...,\sigma_n)$ be a permutation of $(1,...,n)$.
A cyclic order is defined as a permutation modulo cyclic permutations $(\sigma_1,\sigma_2,...,\sigma_n) \rightarrow (\sigma_2,...,\sigma_n,\sigma_1)$.
We may represent a cyclic order by an $n$-gon, where the edges of the $n$-gon are indexed clockwise
by $\sigma_1$, $\sigma_2$, ..., $\sigma_n$.
A dihedral structure is defined as a permutation modulo cyclic permutations and
reflection $(\sigma_1,\sigma_2,...,\sigma_n) \rightarrow (\sigma_n,...,\sigma_2,\sigma_1)$.
We may represent a dihedral structure by an $n$-gon, where the edges of the $n$-gon are indexed either clockwise or anti-clockwise
by $\sigma_1$, $\sigma_2$, ..., $\sigma_n$.
This is illustrated in the left picture of fig.~(\ref{fig_polygon}).

We will need a few definitions related to graphs. In the following we consider tree graphs with $n$ external legs.
For a graph $G$ we denote by $E(G)$ the set of the internal edges and by $s_e$ 
the Lorentz invariant corresponding to the internal edge $e$.

We denote by ${\mathcal T}_n(\sigma)$ the set of all cyclic ordered tree diagrams with trivalent vertices and 
external cyclic order $\sigma$.
The number of graphs in the set ${\mathcal T}_n(\sigma)$ is given by
\bq
 \left| {\mathcal T}_n(\sigma) \right|
 & = &
 \frac{(2n-4)!}{(n-2)!(n-1)!}.
\eq
Two diagrams with different external orders are considered to be equivalent, if we can transform one diagram into the other by a sequence of flips.
Under a flip operation one exchanges at a vertex two branches.
We denote by $\mathrm{CO}(G)$ the set of cyclic orders obtained from the graph $G$ by flipping in all possible ways the branches
at the vertices.
The number of cyclic orders in the set $\mathrm{CO}(G)$ is given by
\bq
 \left| \mathrm{CO}(G) \right| & = & 2^{n-2}.
\eq
We denote by ${\mathcal T}_n(\sigma) \cap {\mathcal T}_n(\tilde{\sigma})$ the set of graphs compatible 
with the external orders $\sigma$ and $\tilde{\sigma}$
and by $n_{\mathrm{flip}}(\sigma,\tilde{\sigma})$ the number of flips needed to transform any graph from
${\mathcal T}_n(\sigma) \cap {\mathcal T}_n(\tilde{\sigma})$ with the external order
$\sigma$ into a graph with the external order $\tilde{\sigma}$.
The number $n_{\mathrm{flip}}(\sigma,\tilde{\sigma})$ will be the same for all graphs from 
${\mathcal T}_n(\sigma) \cap {\mathcal T}_n(\tilde{\sigma})$.

We denote by ${\mathcal U}_n$ the set of all unordered tree graphs with trivalent vertices.
The number of graphs in this set is given by
\bq
 \left| {\mathcal U}_n \right| 
 & = & 
(2n-5)!!
 \;\; = \;\;
 2^{2-n} \frac{\left(2n-4\right)!}{\left(n-2\right)!}.
\eq
Let $G^{\mathrm{unordered}} \in {\mathcal U}_n$ be an unordered tree graph.
We may draw this graph as a cyclic ordered graph in $2^{n-2}$ different ways.
Let $G$ be one possibility. We denote by $\mathrm{co}(G)$ the cyclic order of $G$.
If $G'$ is another possibility of drawing $G^{\mathrm{unordered}}$ as a cyclic ordered graph, the relative sign
between the two graphs is given by
\bq
 \left(-1\right)^{n_{\mathrm{flip}}(\mathrm{co}(G),\mathrm{co}(G'))}.
\eq
The following formula is useful to exchange summation orders:
\bq
\label{exchange_summations}
 \sum\limits_{\sigma \in  S_n/{\mathbb Z}_n}
 \sum\limits_{G \in {\mathcal T}_n(\sigma)}
 & = &
 \sum\limits_{G \in {\mathcal U}_n}
 \sum\limits_{\sigma \in \mathrm{CO}(G)}
\eq
This formula says that it is equivalent either to sum first over all cyclic orders and then within a given cyclic order
over all graphs contributing to it, or to sum first over all unordered graphs and then over all cyclic orders 
compatible with this graph.
In practice the summands will depend on cyclic ordered graphs.
For unordered graphs we will have to pick a cyclic ordered representative and take relative signs between equivalent
representatives into account.

Let us also introduce a notation to specify tree graphs with a fixed cyclic order and trivalent vertices.
We follow ref.~\cite{Tolotti:2013caa}.
Let us assume that the cyclic order is $(1,2,...,n)$.
If we single out one specific external leg (usually we take the last leg $n$),
we speak of a rooted tree, the root being given by the external leg which we singled out.
We may specify a rooted tree by brackets involving the remaining legs, for example
\bq
 \left[ \left[ 1, 2 \right], 3 \right]
\eq
denotes the rooted tree
\bq
\begin{picture}(100,50)(0,25)
\Vertex(50,30){2}
\Vertex(35,45){2}
\Line(50,5)(50,30)
\Line(50,30)(80,60)
\Line(50,30)(20,60)
\Line(35,45)(50,60)
\Text(20,65)[b]{$1$}
\Text(50,65)[b]{$2$}
\Text(80,65)[b]{$3$}
\Text(50,0)[t]{$4$}
\end{picture}.
 \nonumber \\
 \nonumber \\
\eq
In addition we consider non-rooted trees. We define a concatenation operation for two rooted trees:
Let $T_1$ and $T_2$ be two rooted trees with roots $r_1$ and $r_2$.
Then we denote by $(T_1,T_2)$ the non-rooted tree obtained from $T_1$ and $T_2$ by joining the two roots by an edge.
As an example we have:
\bq
\left(
\begin{picture}(60,30)(0,15)
\Vertex(30,20){2}
\Line(30,5)(30,20)
\Line(30,20)(45,35)
\Line(30,20)(15,35)
\Text(15,40)[b]{$1$}
\Text(45,40)[b]{$2$}
\Text(30,0)[t]{$r_{1}$}
\end{picture},
\begin{picture}(60,30)(0,15)
\Vertex(30,20){2}
\Line(30,5)(30,20)
\Line(30,20)(45,35)
\Line(30,20)(15,35)
\Text(15,40)[b]{$3$}
\Text(45,40)[b]{$4$}
\Text(30,0)[t]{$r_{2}$}
\end{picture}
 \right)
 & = &
\begin{picture}(60,30)(-15,15)
\Vertex(15,20){2}
\Vertex(45,20){2}
\Line(15,20)(45,20)
\Line(5,10)(15,20)
\Line(5,30)(15,20)
\Line(55,10)(45,20)
\Line(55,30)(45,20)
\Text(0,10)[r]{$1$}
\Text(0,30)[r]{$2$}
\Text(60,30)[l]{$3$}
\Text(60,10)[l]{$4$}
\end{picture}
 \nonumber \\
\eq
The concatenation operation is symmetric:
\bq
 \left( T_1, T_2 \right) & = & \left( T_2, T_1 \right).
\eq
If $T_1$, $T_2$ and $T_3$ are rooted trees, we have the obvious relations
\bq
 \left( \left[ T_1, T_2 \right], T_3 \right)
 = 
 \left( \left[ T_2, T_3 \right], T_1 \right)
 =
 \left( \left[ T_3, T_1 \right], T_2 \right).
\eq
Every rooted tree can be viewed as a non-rooted tree by simply forgetting that one external leg has been marked as a root.

\subsection{Amplitudes}

In this paper we consider the bi-adjoint scalar amplitudes $m_n(\sigma,\tilde{\sigma},p)$,
the Yang-Mills amplitudes $A_n(\sigma,p,\eps)$ and the graviton amplitudes $M_n(p,\eps,\tilde{\eps})$.
The bi-adjoint scalar amplitude $m_n(\sigma,\tilde{\sigma},p)$ depends on two cyclic orders $\sigma$ and $\tilde{\sigma}$
and arises in the double colour decomposition of the full bi-adjoint tree amplitude:
\bq
m_{n}\left(p\right) 
 & = & 
 \lambda^{n-2} 
 \sum\limits_{\sigma \in S_{n}/Z_{n}} 
 \sum\limits_{\tilde{\sigma} \in S_{n}/Z_{n}} 
 2 \; \mathrm{Tr} \left( T^{a_{\sigma(1)}} ... T^{a_{\sigma(n)}} \right)
 \;\;
 2 \; \mathrm{Tr} \left( \tilde{T}^{b_{\tilde{\sigma}(1)}} ... \tilde{T}^{b_{\tilde{\sigma}(n)}} \right)
 \;\;
 m_{n}\left( \sigma, \tilde{\sigma}, p \right)
 \nonumber \\
\eq
The double-ordered amplitude $m_n(\sigma,\tilde{\sigma},p)$ is rather simple and explicitly given by
\bq
\label{def_m_n}
 m_{n}\left( \sigma, \tilde{\sigma}, p \right)
 & = &
 i \left(-1\right)^{n-3+n_{\mathrm{flip}}(\sigma,\tilde{\sigma})}
 \sum\limits_{G \in {\mathcal T}_n(\sigma) \cap {\mathcal T}_n(\tilde{\sigma})} 
 \;\;\;
 \prod\limits_{e \in E(G)} \frac{1}{s_e}.
\eq
The partial Yang-Mills amplitude $A_n(\sigma,p,\eps)$ appears in the (single) colour decomposition of the
full Yang-Mills tree amplitude:
\bq
\label{colour_decomposition}
{\mathcal A}_{n}\left(p,\eps\right) 
 & = & g^{n-2} \sum\limits_{\sigma \in S_{n}/Z_{n}} 
 2 \; \mathrm{Tr} \left( T^{a_{\sigma(1)}} ... T^{a_{\sigma(n)}} \right)
 \;\;
 A_{n}\left(\sigma,p,\eps\right).
\eq
It is well known that we may eliminate the four-gluon vertex in Yang-Mills theory by introducing an
auxiliary tensor particle \cite{Draggiotis:1998gr,Duhr:2006iq,Weinzierl:2016bus}
with propagator
\bq
\label{tensor_particle_propagator}
 \begin{picture}(100,20)(0,5)
 \Line(20,8)(70,8)
 \Line(20,12)(70,12)
 \Text(15,12)[r]{\footnotesize $[\mu \nu]$}
 \Text(75,12)[l]{\footnotesize $[\rho \sigma]$}
\end{picture} 
& = &
 - \frac{i}{2} \left( g_{\mu\rho} g_{\nu\sigma} - g_{\mu\sigma} g_{\nu\rho} \right),
\eq
and vertex
\bq
\label{tensor_particle_vertex}
\begin{picture}(100,35)(0,40)
\Vertex(50,50){2}
\Line(48,50)(48,20)
\Line(52,50)(52,20)
\Gluon(50,50)(76,65){3}{4}
\Gluon(50,50)(24,65){3}{4}
\Text(22,65)[rc]{$1, \mu$}
\Text(78,65)[lc]{$2, \nu$}
\Text(50,17)[t]{$3, [\rho \sigma]$}
\end{picture}
 & = &
 \frac{i}{\sqrt{2}}
 \left( g^{\mu\rho} g^{\nu\sigma} - g^{\mu\sigma} g^{\nu\rho} \right).
 \\
 \nonumber \\
 \nonumber 
\eq
This allows us to compute $A_n(\sigma,p,\eps)$ from tree diagrams with trivalent vertices only.
Thus we may write
\bq
\label{def_N_Feynman}
 A_n & = &
 i \left(-1\right)^{n-3}
 \sum\limits_{G \in {\mathcal T}_n(\sigma)} N^{\mathrm{Feynman}}(G) \prod\limits_{e \in E(G)} \frac{1}{s_e},
\eq
where the numerators $N^{\mathrm{Feynman}}(G)$ are given by Feynman rules.
Note that the numerators are not BCJ-numerators. Although they do satisfy the anti-symmetry relations, they do in
general not satisfy Jacobi relations.
It is however possible \cite{Bern:2008qj,Bern:2010ue,Bern:2010yg}
to express $A_n$ in a form similar to eq.~(\ref{def_N_Feynman})
\bq
\label{def_N_BCJ}
 A_n & = &
 i \left(-1\right)^{n-3}
 \sum\limits_{G \in {\mathcal T}_n(\sigma)} N^{\mathrm{BCJ}}(G) \prod\limits_{e \in E(G)} \frac{1}{s_e},
\eq
with numerators $N^{\mathrm{BCJ}}(G)$ satisfying anti-symmetry relations and Jacobi relations.
It is further possible to obtain the BCJ-numerators from an effective Lagrangian \cite{Bern:2010yg,Tolotti:2013caa}
\bq
\label{YM_vers2}
 {\mathcal L}_{\mathrm{YM}} + {\mathcal L}_{\mathrm{GF}} & = & 
 \frac{1}{2g^2} \sum\limits_{n=2}^\infty {\mathcal L}^{(n)},
\eq
where ${\mathcal L}^{(n)}$ contains $n$ fields.
The terms ${\mathcal L}^{(2)}$, ${\mathcal L}^{(3)}$ and ${\mathcal L}^{(4)}$ agree with the standard terms.
Let us introduce the Lie-algebra valued field
\bq
 {\bf A}_\mu & = & \frac{g}{i} T^a A^a_\mu.
\eq
Then the terms ${\mathcal L}^{(2)}$, ${\mathcal L}^{(3)}$ and ${\mathcal L}^{(4)}$ are given by
\bq
 {\mathcal L}^{(2)} & = &
 -2 \mbox{Tr} \; {\bf A}_{\mu} \Box {\bf A}^{\mu},
 \nonumber \\
 {\mathcal L}^{(3)} & = &
 4 \mbox{Tr} \; \left( \partial_\mu {\bf A}_{\nu} \right) \left[ {\bf A}^{\mu}, {\bf A}^{\nu} \right],
 \nonumber \\
 {\mathcal L}^{(4)} & = &
 - g^{\mu_1\mu_3} g^{\mu_2\mu_4} g_{\nu_1\nu_2}
 \frac{\partial^{\nu_1}_{12} \partial^{\nu_2}_{34} }{\Box_{12}} 
 \; \mbox{Tr} \; \left[ {\bf A}_{\mu_1}, {\bf A}_{\mu_2} \right] \left[ {\bf A}_{\mu_3}, {\bf A}_{\mu_4} \right].
\eq
The subscripts on the derivatives indicate on which fields they act.
For ${\mathcal L}^{(4)}$ we have introduced a factor $1/\Box_{12}$, which assigns an intermediate propagator to the four-gluon vertex.
Note that this factor cancels against the factor  $(-g_{\nu_1\nu_2}\partial^{\nu_1}_{12} \partial^{\nu_2}_{34}) = \Box_{12}$.
Leaving the two factors uncancelled keeps the information on the assignment of terms to diagrams with three-valent vertices only.
This corresponds exactly to the introduction of an auxiliary particle through eq.~(\ref{tensor_particle_propagator}) and eq.~(\ref{tensor_particle_vertex}).

The terms ${\mathcal L}^{(n)}$ for $n\ge 5$ are equivalent to zero.
They ensure that BCJ-numerators are obtained from the Feynman rules.
To give an example we may take for ${\mathcal L}^{(5)}$
\bq
\label{example_L_5}
 {\mathcal L}^{(5)} & = &
  4 g^{\mu_1\mu_3} g^{\mu_2\mu_4} \frac{\partial_{1}^{\mu_5}}{\Box_{123}}
 \left(
 \mbox{Tr} \;  \left[ \left[ \left[ {\bf A}_{\mu_1}, {\bf A}_{\mu_2} \right], {\bf A}_{\mu_3} \right], {\bf A}_{\mu_4} \right] {\bf A}_{\mu_5}
 \right.
 \nonumber \\
 & &
 \left.
  + 
 \mbox{Tr} \; \left[ \left[ {\bf A}_{\mu_3}, {\bf A}_{\mu_4} \right], \left[ {\bf A}_{\mu_1}, {\bf A}_{\mu_2} \right] \right] {\bf A}_{\mu_5}
  + 
 \mbox{Tr} \; \left[ \left[ {\bf A}_{\mu_4}, \left[ {\bf A}_{\mu_1}, {\bf A}_{\mu_2} \right] \right], {\bf A}_{\mu_3} \right] {\bf A}_{\mu_5}
 \right).
\eq
The term ${\mathcal L}^{(5)}$ is equal to zero due to the Jacobi identity
involving the expressions $[{\bf A}_{\mu_1},{\bf A}_{\mu_2}]$, ${\bf A}_{\mu_3}$ and ${\bf A}_{\mu_4}$.
However, the term ${\mathcal L}^{(5)}$ generates a five-valent vertex.
This five-valent vertex gives a non-vanishing contribution to individual numerators.
In a partial amplitude the sum of all terms related to the five-valent vertex adds up to zero.
We note that the terms ${\mathcal L}^{(n)}$ for $n\ge 5$ are not unique.
For example, we may replace in eq.~(\ref{example_L_5})
\bq
\label{freedom_five}
 4 g^{\mu_1\mu_3} g^{\mu_2\mu_4} \partial_{1}^{\mu_5}
 & \rightarrow &
 4 g^{\mu_1\mu_3} g^{\mu_2\mu_4} \partial_{1}^{\mu_5}
 -
 4 a \left( g^{\mu_1\mu_3} g^{\mu_2\mu_4} \partial_{1}^{\mu_5} - g^{\mu_1\mu_3} g^{\mu_4\mu_5} \partial_{4}^{\mu_2} \right),
\eq
where $a$ is a free parameter.

From the effective Lagrangian in eq.~(\ref{YM_vers2}) we obtain
BCJ-numerators, which are polynomials in $p$ and $\eps$.
The BCJ-numerators $N^{\mathrm{BCJ}}(G)$
inherit their graph structure from the underlying graph $G$.
Let $I$ be a subset of $\{1,2,...,n\}$. We say that $N^{\mathrm{BCJ}}(G)$ factorises in the channel $I$, if there is an edge $e \in G$, 
such that $s_I=s_e$, otherwise we say that $N^{\mathrm{BCJ}}(G)$ does not factorise in the channel $I$.
In other words, the factorisation channels of a graph $G$ correspond exactly to the internal edges (or propagators) of the graph.

Combining the anti-symmetry of the vertices and the Jacobi identity one has
\bq
\label{STU_relation}
\begin{picture}(110,30)(0,30)
\Line(10,10)(90,10)
\Vertex(50,30){2}
\Vertex(50,10){2}
\Line(50,10)(50,30)
\Line(50,30)(70,50)
\Line(50,30)(30,50)
\Text(5,10)[r]{$1$}
\Text(30,55)[b]{$2$}
\Text(70,55)[b]{$3$}
\Text(95,10)[l]{$4$}
\end{picture}
 & = & 
\begin{picture}(120,30)(-10,30)
\Line(10,10)(90,10)
\Vertex(35,10){2}
\Vertex(65,10){2}
\Line(35,10)(35,40)
\Line(65,10)(65,40)
\Text(5,10)[r]{$1$}
\Text(35,45)[b]{$2$}
\Text(65,45)[b]{$3$}
\Text(95,10)[l]{$4$}
\end{picture}
-
\begin{picture}(110,30)(-10,30)
\Line(10,10)(90,10)
\Vertex(35,10){2}
\Vertex(65,10){2}
\Line(35,10)(35,40)
\Line(65,10)(65,40)
\Text(5,10)[r]{$1$}
\Text(35,45)[b]{$3$}
\Text(65,45)[b]{$2$}
\Text(95,10)[l]{$4$}
\end{picture}.
 \\
 \nonumber \\
 \nonumber
\eq
Eq.~(\ref{STU_relation}) is called a STU-relation.
With the help of eq.~(\ref{STU_relation}) we may reduce any tree graph with $n$ external legs
and containing only trivalent vertices to a multi-peripheral form with respect to $1$ and $n$.
We say that a graph is multi-peripheral (or in comb form) with respect to $1$ and $n$,
if all other external legs connect directly to the line from $1$ to $n$,
i.e. there are no non-trivial sub-trees attached to this line.
A graph in multi-peripheral form can be drawn as
\bq
\begin{picture}(210,60)(0,0)
\Line(10,10)(200,10)
\Vertex(35,10){2}
\Vertex(65,10){2}
\Vertex(95,10){2}
\Vertex(145,10){2}
\Vertex(175,10){2}
\Line(35,10)(35,40)
\Line(65,10)(65,40)
\Line(95,10)(95,40)
\Line(145,10)(145,40)
\Line(175,10)(175,40)
\Text(5,10)[r]{$1$}
\Text(35,45)[b]{$\alpha_2$}
\Text(65,45)[b]{$\alpha_3$}
\Text(120,25)[b]{$...$}
\Text(175,45)[b]{$\alpha_{n-1}$}
\Text(205,10)[l]{$n$.}
\end{picture}
\eq
Let us set $\kappa=(1,\alpha,n)$.
We denote the BCJ-numerator of a multi-peripheral graph by
\bq
\label{def_BCJ_numerator}
 N^{\mathrm{BCJ}}_{\mathrm{comb}}\left(\kappa\right).
\eq
There are $(n-2)!$ multi-peripheral BCJ-numerators, indexed by a permutation $(\alpha_2,...,\alpha_{n-1})$ of $(2,...,n-1)$.

After this excursion towards BCJ-numerators let us turn back to scattering
amplitudes.
The graviton amplitude $M_n(p,\eps,\tilde{\eps})$ is obtained by expanding the Einstein-Hilbert action
\bq
 {\mathcal L}_{\mathrm{EH}}
 & = &
 - \frac{2}{\kappa^2} 
 \sqrt{-g} R
\eq
around the flat Minkowski metric
\bq
 g_{\mu\nu} & = &
 \eta_{\mu\nu} + \kappa h_{\mu\nu}.
\eq
$h_{\mu\nu}$ is the graviton field. The polarisation of a graviton is described by a product
of two spin-$1$ polarisation vectors:
\bq
 \mbox{either} \;\;\; \eps_{\mu}^+ \tilde{\eps}_{\nu}^+ \;\;\;
 \mbox{or} \;\;\; \eps_{\mu}^- \tilde{\eps}_{\nu}^-.
\eq
For $n$ external gravitons we collect the first polarisation vector in $\eps$, and the second 
polarisation vector in $\tilde{\eps}$.

\subsection{The scattering equations}

Let us consider $z=(z_1,z_2,...,z_n) \in \left( {\mathbb C} {\mathbb P}^1 \right)^n$
with $z_i \neq z_j$.
One defines
for $1\le i \le n$
\bq
 f_i\left(z,p\right) & = & 
 \sum\limits_{j=1, j \neq i}^n \frac{ 2 p_i \cdot p_j}{z_i - z_j}.
\eq
The scattering equations are given by \cite{Cachazo:2013gna,Cachazo:2013hca,Cachazo:2013iea}
\bq
\label{scattering_equations}
 f_i\left(z,p\right) & = & 0,
 \;\;\;\;\;\;
 i \in \{1,...,n\}.
\eq
For fixed momenta $p$ a solution of the scattering equations is a point $z \in \left( {\mathbb C} {\mathbb P}^1 \right)^n$,
such that the scattering equations in eq.~(\ref{scattering_equations}) are satisfied.

The scattering equations are invariant under the projective special linear group
$\mathrm{PSL}(2,{\mathbb C})=\mathrm{SL}(2,{\mathbb C})/{\mathbb Z}_2$.
Here, ${\mathbb Z}_2$ is given by $\{ {\bf 1}, -{\bf 1} \}$, with ${\bf 1}$ denoting
the unit matrix.
The matrix 
\bq
 g = \left(\begin{array}{cc} a & b \\ c & d \\ \end{array} \right) & \in & \mathrm{PSL}(2,{\mathbb C})
\eq
acts on all $z_j$ by
\bq
 g \cdot z_i & = & \frac{a z_i + b}{c z_i +d},
 \;\;\;\;\;\;
 i \in \{1,...,n\}.
\eq
We write $g \cdot(z_1,...,z_n) = (g \cdot z_1, ..., g \cdot z_n)$.
If $(z_1,z_2, ..., z_n)$ is a solution of eq.~(\ref{scattering_equations}),
then also $(z_1',z_2', ..., z_n') = g \cdot (z_1,z_2, ..., z_n)$ is a solution.
Two solutions which are related by a $\mathrm{PSL}(2,{\mathbb C})$-transformation are called equivalent solutions.
From B\'ezout's theorem \cite{Griffiths:book}
it follows that there are $(n-3)!$ inequivalent solutions of the scattering equations 
not related by a $\mathrm{PSL}(2,{\mathbb C})$-trans\-for\-ma\-tion.

Following \cite{Mizera:2017rqa} one defines a one-form $\eta$ by
\bq
\label{def_eta}
 \eta & = &
 \sum\limits_{i=1}^n f_i\left(z,p\right) dz_i.
\eq
The one-form $\eta$ defines the twist, such that amplitudes are expressed
as twisted intersection numbers of two cocycles.
The two cocycles are differential $(n-3)$-forms, and their properties are the subject of this paper.

\subsection{The CHY representation of amplitudes}

There are a large variety of theories, in which the Born amplitudes have a CHY representation, i.e.
the Born amplitudes can be written in the form
\bq
\label{CHY_int_representation_A_n}
 A_n
 & = &
 i \oint\limits_{\mathcal C} d\Omega_{\mathrm{CHY}} \; F\left(z\right).
\eq
The measure $d\Omega_{\mathrm{CHY}}$ is defined by
\bq
 d\Omega_{\mathrm{CHY}}
 & = &
 \frac{1}{\left(2\pi i\right)^{n-3}}
 \frac{d^nz}{d\omega}
 \;
 \prod{}' \frac{1}{f_a\left(z,p\right)},
\eq
where
\bq
 \prod{}' \frac{1}{f_a\left(z,p\right)}
 & = & 
 \left(-1\right)^{i+j+k}
 \left( z_i - z_j \right) \left( z_j - z_k \right) \left( z_k - z_i \right)
 \prod\limits_{a \neq i,j,k} \frac{1}{f_a\left(z,p\right)},
\eq
and
\bq
 d\omega
 & = &
 \left(-1\right)^{p+q+r}
 \frac{dz_p dz_q dz_r}{\left( z_p - z_q \right) \left( z_q - z_r \right) \left( z_r - z_q \right)}.
\eq
The primed product of $1/f_a$ is independent of the choice of $i$, $j$, $k$ and
takes into account that only $(n-3)$ of the scattering equations are independent.
The quantity $d\omega$ is independent of the choice of $p$, $q$, $r$
and corresponds to the invariant measure on $\mathrm{PSL}(2,{\mathbb C})$.
The integration contour ${\mathcal C}$ encircles the inequivalent solutions of the scattering equations.
Under $\mathrm{PSL}(2,{\mathbb C})$-transformations the integrand $F(z)$ transforms as
\bq
 F\left( g \cdot z \right) 
 & = &
 \left( \prod\limits_{j=1}^n \left( c z_j +d \right)^4 \right)
 F\left( z \right).
\eq
It is often the case that $F(z)$ factors
\bq
 F\left(z\right)
 & = &
 F_1\left(z\right) F_2\left(z\right),
\eq
where each factor transforms as
\bq
 F_i\left( g \cdot z \right) 
 & = &
 \left( \prod\limits_{j=1}^n \left( c z_j +d \right)^2 \right)
 F_i\left( z \right),
 \;\;\;\;\;\; i \in \{1,2\}.
\eq
We call the factors $F_1(z)$ and $F_2(z)$ half-integrands.
In this paper we consider the standard examples of amplitudes in a bi-adjoint scalar theory, Yang-Mills theory and gravity.
For these amplitudes the CHY integrand is constructed from two building blocks, a cyclic factor $C(\sigma,z)$ defined by
\bq
\label{def_cyclic_factor}
 C\left(\sigma,z\right)
 & = & 
 \frac{1}{\left(z_{\sigma_1} - z_{\sigma_2} \right) \left( z_{\sigma_2} - z_{\sigma_3} \right) ... \left( z_{\sigma_n} - z_{\sigma_1} \right)},
\eq
and
a polarisation factor $E(p,\eps,z)$.
The original definition of the polarisation factor in terms of a reduced Pfaffian is given in 
appendix~\ref{sect:CHY_integrands}.
The CHY integrands are then given by
\bq
 F_{\mathrm{bi-adjoint}}
 & = &
 C\left(\sigma,z\right) C\left(\tilde{\sigma},z\right),
 \nonumber \\
 F_{\mathrm{YM}}
 & = &
 C\left(\sigma,z\right) E\left(p,\eps,z\right),
 \nonumber \\
 F_{\mathrm{gravity}}
 & = &
 E\left(p,\eps,z\right) E\left(p,\tilde{\eps},z\right).
\eq
Each half-integrand defines a $(n-3)$-form
\bq
 \Omega_{\mathrm{scattering}}^{\mathrm{cyclic}}\left(\sigma,z\right) 
 & = &
  C\left(\sigma,z\right) \frac{d^nz}{d\omega},
 \nonumber \\
 \Omega_{\mathrm{scattering}}^{\mathrm{pol}}\left(p,\eps,z\right)
 & = &
  E\left(p,\eps,z\right) \frac{d^nz}{d\omega}.
\eq
Mizera \cite{Mizera:2017rqa} has shown recently that the amplitudes are given as twisted intersection
numbers of these forms, twisted by the one-form $\eta$, i.e.
\bq
\label{amplitudes_twsited_intersection}
 m_{n}\left( \sigma, \tilde{\sigma}, p \right)
 & = &
 i \left( \Omega_{\mathrm{scattering}}^{\mathrm{cyclic}}\left(\sigma,z\right), \Omega_{\mathrm{scattering}}^{\mathrm{cyclic}}\left(\tilde{\sigma},z\right) \right)_\eta,
 \nonumber \\
 A_n(\sigma,p,\eps)
 & = &
 i \left( \Omega_{\mathrm{scattering}}^{\mathrm{cyclic}}\left(\sigma,z\right), \Omega_{\mathrm{scattering}}^{\mathrm{pol}}\left(p,\eps,z\right) \right)_\eta,
 \nonumber \\
 M_n(p,\eps,\tilde{\eps})
 & = &
 i \left( \Omega_{\mathrm{scattering}}^{\mathrm{pol}}\left(p,\eps,z\right), \Omega_{\mathrm{scattering}}^{\mathrm{pol}}\left(p,\tilde{\eps},z\right) \right)_\eta,
\eq
where $(A,B)_\eta$ denotes the intersection number of two cocycles twisted by $\eta$.
Eq.~(\ref{amplitudes_twsited_intersection}) depends only on the values of the scattering forms on the sub-variety defined by the scattering
equations.
However, the scattering forms $\Omega_{\mathrm{scattering}}^{\mathrm{cyclic}}$ and $\Omega_{\mathrm{scattering}}^{\mathrm{pol}}$
themselves
do not know about the scattering equations.
The scattering equations enter only through the twist $\eta$.
It turns out that the cyclic scattering form $\Omega_{\mathrm{scattering}}^{\mathrm{cyclic}}$ has in addition nice mathematical
properties away from the zero-dimensional sub-variety defined by the scattering equations.
In this paper we study these properties. 
We also show that with a suitable re-definition of the polarisation factor $E(p,\eps,z)$ the same properties hold
for the scattering form $\Omega_{\mathrm{scattering}}^{\mathrm{pol}}$.

\subsection{Multivariate residues}

In this paragraph we review multivariate residues of differential forms.
We follow ref.~\cite{Abreu:2017ptx}.
Let $X$ be a $n$-dimensional variety and $Y$ a co-dimension one sub-variety.
Let us choose a coordinate system such that $Y$ is given locally by $z_1=0$.
Assume that $\Omega$ has a pole of order $k$ on $Y$. Then $\Omega$ may be written as
\bq
 \Omega & = &
 \frac{dz_1}{z_1^k} \wedge \psi + \theta,
\eq
where the $(n-1)$-form $\psi$ is regular and non-vanishing on $Y$, and the $n$-form $\theta$ has at most
a pole of order $(k-1)$ on $Y$.
We may reduce poles of order $k>1$ to poles of order $1$ and exact forms due to the identity
\bq
 \frac{dz_1}{z_1^k} \wedge \psi + \theta
 & = &
 d \left(-\frac{\psi}{\left(k-1\right)z_1^{k-1}} \right)
 + \frac{d\psi}{\left(k-1\right) z_1^{k-1}} + \theta.
\eq
Thus every form $\Omega$ is equivalent (up to an exact form) to a form $\Omega_1$ with at most a single pole
on $Y$.
For
\bq
 \Omega_1 & = &
 \frac{dz_1}{z_1} \wedge \psi_1 + \theta_1,
\eq
we set
\bq
 \mathrm{Res}_Y\left(\Omega_1\right)
 & = &
 \left. \psi_1 \right|_Y,
\eq
and if $\Omega_1$ is equivalent to $\Omega$ up to an exact form
\bq
 \mathrm{Res}_Y\left(\Omega\right)
 & = &
 \mathrm{Res}_Y\left(\Omega_1\right).
\eq
Multivariate residues are defined as follows: Suppose we have two co-dimension one sub-varieties $Y_1$ and $Y_2$ defined by
$z_1=0$ and $z_2=0$, respectively.
Again we may reduce higher poles to simple poles modulo exact forms.
Let us therefore consider
\bq
 \Omega & = &
 \frac{dz_1}{z_1} \wedge \frac{dz_2}{z_2} \wedge \psi_{12}
 + \frac{dz_1}{z_1} \wedge \psi_1
 + \frac{dz_2}{z_2} \wedge \psi_2
 + \theta,
\eq
where $\psi_{12}$ is regular on $Y_1 \cap Y_2$, $\psi_j$ is regular on $Y_j$ and $\theta$ is regular 
on $Y_1 \cup Y_2$.
One sets
\bq
 \mathrm{Res}_{Y_1,Y_2}\left(\Omega\right)
 & = &
 \left. \psi_{12} \right|_{Y_1 \cap Y_2}.
\eq
Note that the residue is anti-symmetric with respect to the order of the hypersurfaces:
\bq
 \mathrm{Res}_{Y_2,Y_1}\left(\Omega\right)
 & = &
 -
 \mathrm{Res}_{Y_1,Y_2}\left(\Omega\right).
\eq
Multivariate residues for several co-dimension one sub-varieties $Y_1$, ..., $Y_m$ are defined analogously.

\section{The moduli space of genus $0$ curves with $n$ distinct marked points}
\label{sect:M_0_n}

\begin{figure}
\begin{center}
\includegraphics[scale=0.8]{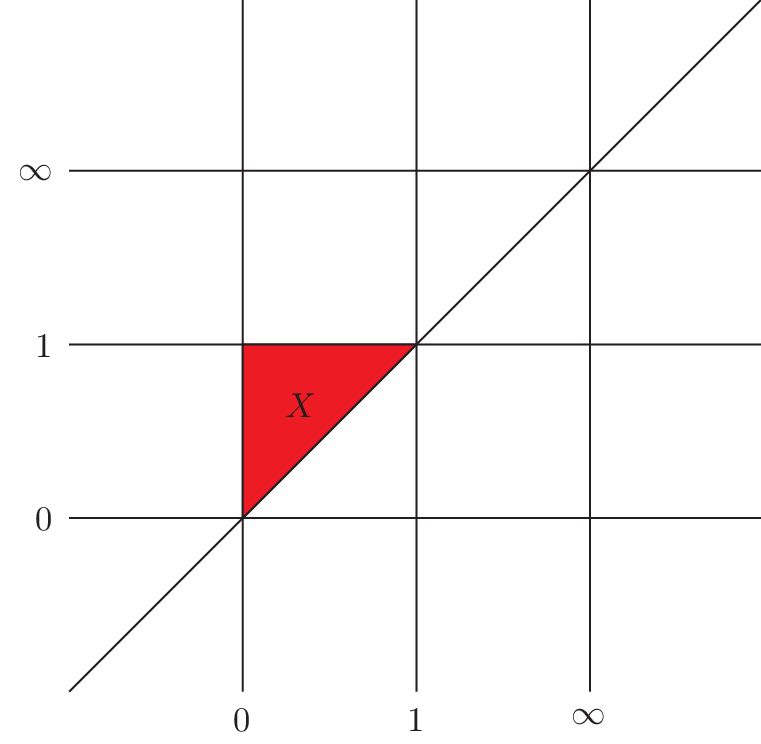}
\hspace*{10mm}
\includegraphics[scale=0.8]{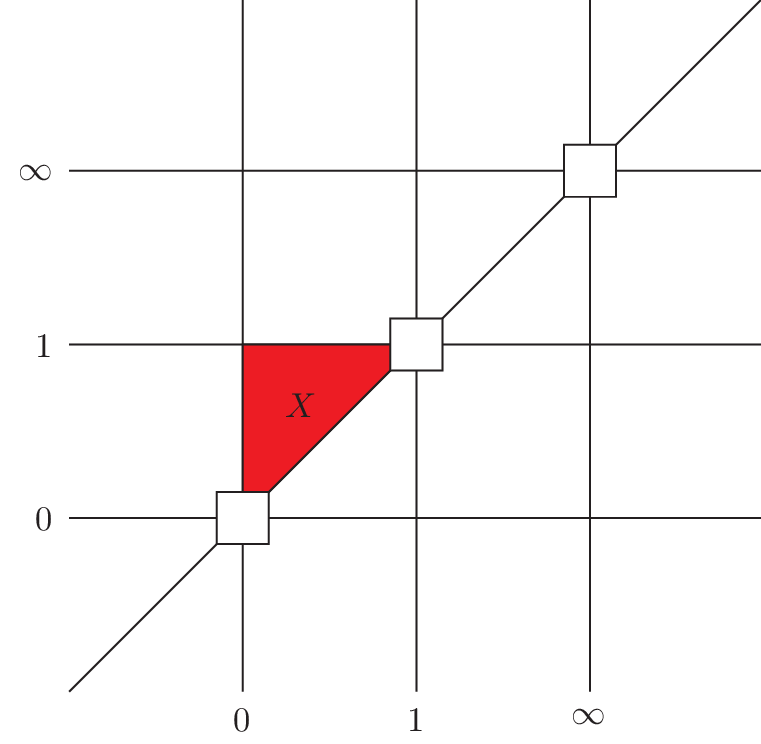}
\end{center}
\caption{
The moduli space ${\mathcal M}_{0,5}({\mathbb R})$ (left).
The region $X$ is bounded by $z_2=0$, $z_3=1$ and $z_2=z_3$.
The right figure shows $\overline{\mathcal M}_{0,5}({\mathbb R})$, obtained from ${\mathcal M}_{0,5}({\mathbb R})$ by blowing up the points
$(z_2,z_3)=(0,0)$, $(z_2,z_3)=(1,1)$ and $(z_2,z_3)=(\infty,\infty)$.
}
\label{fig_M_0_5}
\end{figure}

Let us consider a Riemann sphere (i.e. an algebraic curve of genus zero) with $n$ distinct marked points.
The moduli space of genus $0$ curves with $n$ distinct marked points is denoted
by
\bq
 {\mathcal M}_{0,n}
 & = &
 \left\{ z \in \left( {\mathbb C} {\mathbb P}^1 \right)^n : z_i \neq z_j \right\}/\mathrm{PSL}\left(2,{\mathbb C}\right).
\eq
${\mathcal M}_{0,n}$ is an affine algebraic variety of dimension $(n-3)$.
We may use the freedom of $\mathrm{PSL}(2,{\mathbb C})$-transformations to fix three points.
The standard choice will be $z_1=0$, $z_{n-1}=1$ and $z_n=\infty$.
Thus
\bq
 {\mathcal M}_{0,n}
 & = &
 \left\{ (z_2,...,z_{n-2}) \in {\mathbb C}^{n-3} \; : \; z_i \neq z_j, \; z_i \neq 0, \; z_i \neq 1 \right\}.
\eq
We denote the set of real points by ${\mathcal M}_{0,n}({\mathbb R})$:
\bq
 {\mathcal M}_{0,n}\left({\mathbb R}\right)
 & = &
 \left\{ (z_2,...,z_{n-2}) \in {\mathbb R}^{n-3} \; : \; z_i \neq z_j, \; z_i \neq 0, \; z_i \neq 1 \right\}.
\eq
In fig.~(\ref{fig_M_0_5}) we sketch the moduli space ${\mathcal M}_{0,5}({\mathbb R})$.
In this example the region $X$ is bounded by $z_2=0$, $z_3=1$ and $z_2=z_3$.
In general there will be points, where the boundaries do not cross normally.
For the region $X$ in the example above this occurs for $(z_2,z_3)=(0,0)$ and $(z_2,z_3)=(1,1)$. 
We denote by $\overline{\mathcal M}_{0,n}$ the blow-up of ${\mathcal M}_{0,n}$ in all those points, such that
in $\overline{\mathcal M}_{0,n}$ all boundaries cross normally.
In this way the region $X$ of our example transforms from a triangle in ${\mathcal M}_{0,5}({\mathbb R})$ into
a pentagon in $\overline{\mathcal M}_{0,5}({\mathbb R})$.

\subsection{The Deligne-Mumford-Knudsen compactification}

Let us now review a systematic way to construct $\overline{\mathcal M}_{0,n}$.
There is a smooth compactification 
\bq
 \mathcal M_{0,n} \subset \overline{\mathcal M}_{0,n},
\eq
known as the Deligne-Mumford-Knudsen compactification \cite{Deligne:1969,Knudsen:1976,Knudsen:1983,Knudsen:1983a},
such that $\overline{\mathcal M}_{0,n} \backslash \mathcal M_{0,n}$ is a smooth normal
crossing divisor.
In order to describe $\overline{\mathcal M}_{0,n}$ we follow ref.~\cite{Brown:2006}.
The construction proceeds through intermediate spaces $\mathcal M_{0,n}^\pi$, labelled by a dihedral structure $\pi$,
such that
\bq
 \mathcal M_{0,n} \subset \mathcal M_{0,n}^\pi \subset \overline{\mathcal M}_{0,n}.
\eq
Let $z=(z_1,...,z_n)$ denote the set of the $n$ marked points on the curve.
In the following we will use the notation 
\bq
 \mathcal M_{0,z}
\eq
for $\mathcal M_{0,n}$.
This notation allows us to distinguish $\mathcal M_{0,z'}$ from $\mathcal M_{0,z''}$ if $z'$ and $z''$ are two non-identical subsets of $z$
with $k$ elements each (i.e. $z' \neq z''$ but $|z'|=|z''|=k$).
Let $\pi$ denote a permutation of $(1,...,n)$, which defines a dihedral structure.
We may draw a regular $n$-gon, where the edges are labelled by $z_{\pi_1}$, $z_{\pi_2}$, ..., $z_{\pi_n}$ in this order.
In order to keep the notation simple let us assume that $\pi=(1,2,...,n)$. Then the edges are labelled by $z_1$, $z_2$, ..., $z_n$.
A chord of the polygon connects two non-adjacent vertices and may be specified by giving the two edges preceding the two vertices in
the clockwise orientation.
Thus $(i,j)$ denotes the chord from the vertex between edge $z_i$ and $z_{i+1}$ to the vertex between the edge $z_j$ and $z_{j+1}$.
There are
\bq
 \frac{1}{2} n \left( n-3 \right)
\eq
chords for a regular $n$-gon.
We denote by $\chi(z,\pi)$ the set of all chords of the $n$-gon defined by the set $z$ and the dihedral structure $\pi$.
\begin{figure}
\begin{center}
\includegraphics[scale=0.6]{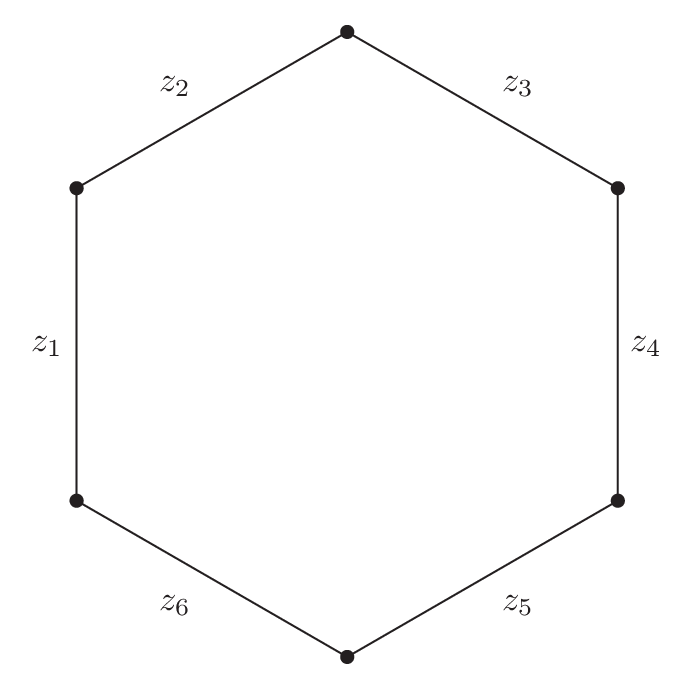}
\hspace*{10mm}
\includegraphics[scale=0.6]{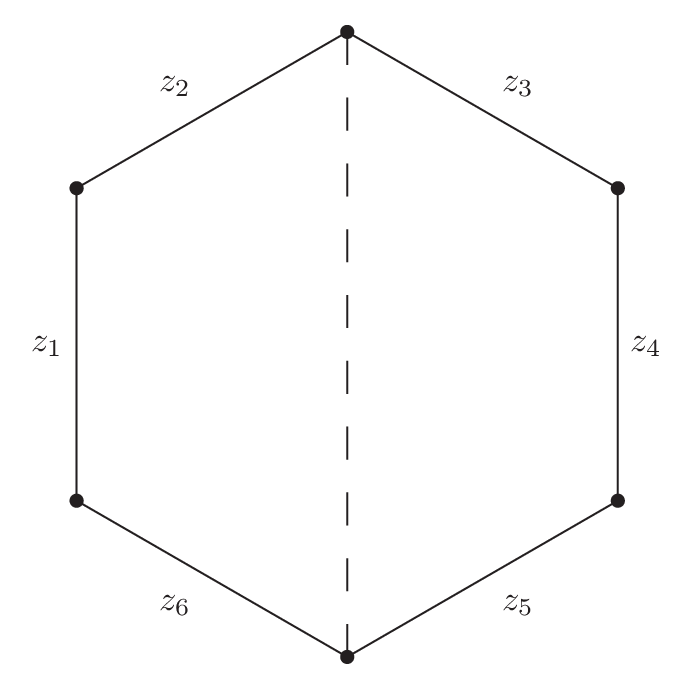}
\hspace*{10mm}
\includegraphics[scale=0.6]{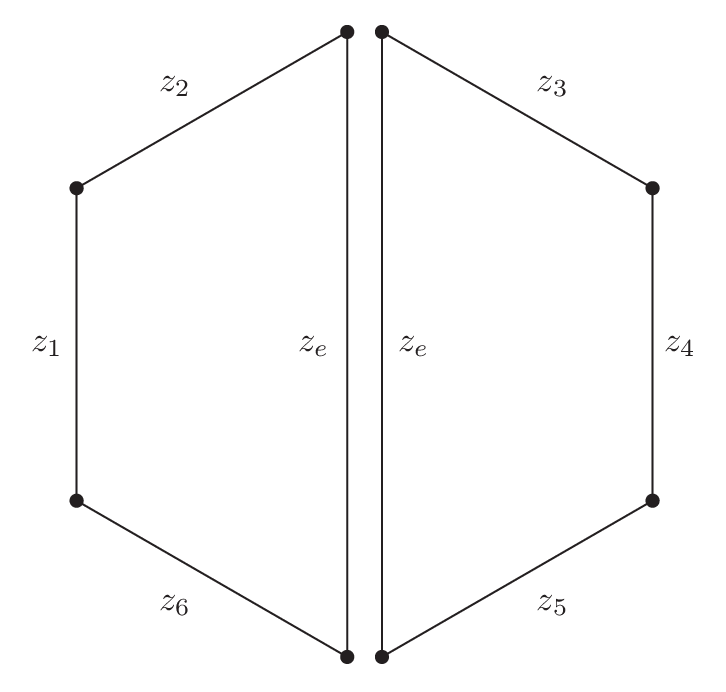}
\end{center}
\caption{
A hexagon, where the edges are labelled by the cyclic ordered variables $(z_1,z_2,...,z_6)$ (left picture).
The middle picture shows the chord $(2,5)$.
Right picture: A chord divides the hexagon into two lower $n$-gons, in this case two quadrangles.
}
\label{fig_polygon}
\end{figure}
Each chord defines a cross-ratio as follows:
\bq
 u_{i,j} & = &
 \frac{\left(z_i-z_{j+1}\right) \left(z_{i+1}-z_j\right)}{\left(z_i-z_j\right)\left(z_{i+1}-z_{j+1}\right)}.
\eq
The cross-ratio is invariant under 
$\mathrm{PSL}(2,{\mathbb C})$-transformations.
Each cross-ratio defines a function
\bq
 \mathcal M_{0,z}
 & \rightarrow &
 {\mathbb C} {\mathbb P}^1 \backslash \{0,1,\infty\},
\eq
or equivalently
\bq
 \mathcal M_{0,z}
 & \rightarrow &
 {\mathbb C} \backslash \{0,1\}.
\eq
The set of all cross-ratios for a given dihedral structure $\pi$ defines an embedding
\bq
 \mathcal M_{0,z}
 & \rightarrow &
 {\mathbb C}^{n (n-3)/2}.
\eq
One defines the dihedral extension $\mathcal M_{0,z}^\pi$ of $\mathcal M_{0,z}$ to be the Zariski closure of the image of this embedding.
The Deligne-Mumford-Knudsen compactification is obtained by gluing these charts together:
\bq
 \overline{\mathcal M}_{0,z}
 & = &
 \bigcup\limits_{\pi} \; \mathcal M_{0,z}^\pi,
\eq
where $\pi$ ranges over the $(n-1)!/2$ inequivalent dihedral structures.

\subsection{The dihedral extension}

Central to our study will be the dihedral extension $\mathcal M_{0,z}^\pi$.
We recall that the construction of $\mathcal M_{0,z}^\pi$ requires the specification
of a dihedral structure $\pi$ (i.e. a permutation up to cyclic permutations and reflection).
We will need a few properties of the dihedral extension $\mathcal M_{0,z}^\pi$ \cite{Brown:2006}:
\begin{enumerate}
\item The complement $\mathcal M_{0,z}^\pi \backslash \mathcal M_{0,z}$ is a normal crossing divisor, whose irreducible components are
\bq
\label{def_divisors}
 D_{i j} & = & \left\{ \; u_{i,j} \; = \; 0 \; \right\},
\eq
indexed by the chords $(i,j) \in \chi(z,\pi)$. 
\item Each divisor is again a product of spaces of the same type:
Let us consider a chord $(i,j)$. This chord decomposes the original polygon $(z,\pi)$ into two smaller polygons,
as shown in fig.~\ref{fig_polygon}.
We denote the new edge by $z_e$.
The set of edges for the two smaller polygons are
$z' \cup \{z_e\}$ and $z'' \cup \{z_e\}$, where $z = z' \cup z''$ and $z' \cap z'' = \emptyset$.
The two smaller polygons inherit their dihedral structures $\pi'$ and $\pi''$ from $\pi$ and the chord $(i,j)$.
We have
\bq
\label{product_structure}
 D_{i j} 
 & \cong &
 \mathcal M_{0,z' \cup \{z_e\}}^{\pi'}
 \times
 \mathcal M_{0,z'' \cup \{z_e\}}^{\pi''}.
\eq
\end{enumerate}

\subsection{The associahedron}

Let us now consider the space of real points.
For a given set $z$ and dihedral structure $\pi$ we set
\bq
 X_{0,z}^\pi
 & = &
 \left\{
  u_{i,j} > 0 \; : \; (i,j) \in \chi(z,\pi)
 \right\}
\eq
and
\bq
 \overline{X}_{0,z}^\pi
 & = &
 \left\{
  u_{i,j} \ge 0 \; : \; (i,j) \in \chi(z,\pi)
 \right\}.
\eq
One has 
\bq
 \mathcal M_{0,n}({\mathbb R})
 & = &
 \bigsqcup\limits_{\pi} \; X_{0,z}^\pi,
\eq
where 
$\pi$ ranges again over the $(n-1)!/2$ inequivalent dihedral structures.

For a given set $z$ and dihedral structure $\pi$ the cell $\overline{X}_{0,z}^\pi$ is called 
a Stasheff polytope or associahedron \cite{Stasheff:1963a,Stasheff:1963b,Devadoss:1998,Devadoss:2004}.
The associahedron has the properties
\begin{enumerate}
\item Its facets (i.e. codimension one faces) 
\bq
 F_{i j} & = & \left\{ \; u_{i,j} \; = \; 0 \; \right\},
\eq
are indexed by the chords 
$(i,j) \in \chi(z,\pi)$.
\item From eq.~(\ref{product_structure}) it follows that each facet is a product
\bq
 F_{ij} & = &
 \overline{X}_{0,z' \cup \{z_e\}}^{\pi'}
 \times
 \overline{X}_{0,z'' \cup \{z_e\}}^{\pi''}.
\eq
\item Two facets $F_{i j}$ and $F_{k l}$ meet if and only if the chords $(i,j)$ and $(k,l)$ do not cross.
\item Faces of codimension $k$ are given by sets of $k$ non-crossing chords.
In particular, the set of vertices of $\overline{X}_{0,z}^\pi$ are in one-to-one
correspondence with the set of triangulations of the $n$-gon defined by the set $z$ and the
dihedral structure $\pi$.
\end{enumerate}
Properties (1) and (2) are the analogues of eq.~(\ref{def_divisors}) 
and eq.~(\ref{product_structure}), respectively.

\subsection{Coordinates on $\mathcal M_{0,z}^\pi$}

Let us now fix a dihedral structure $\pi$. Without loss of generality we may take the cyclic order to be
$(1,2,...,n)$.
Let us consider a chord from $\chi(z,\pi)$. Due to cyclic invariance we may limit ourselves to chords
of the form $(i,n)$.
With the gauge choice $z_1=0$, $z_{n-1}=1$ and $z_n=\infty$ we have
\bq
 u_{2,n} \;\; = \;\; \frac{z_2}{z_3},
 \;\;\;\;\;\;
 ...
 \;\;\;\;\;\;
 u_{(n-3),n} \;\; = \;\; \frac{z_{n-3}}{z_{n-2}},
 \;\;\;\;\;\;
 u_{(n-2),n} \;\; = \;\; z_{n-2},
\eq
and hence
\bq
 z_i & = &
 \prod\limits_{j=i}^{n-2} u_{j,n},
 \;\;\;\;\;\;\;\;\;\;\;\;
 i \in \{2,...,n-2\}.
\eq
Thus we may use as coordinates on $\mathcal M_{0,z}^\pi$ instead of the $(n-3)$ coordinates
$(z_2,...,z_{n-2})$ the $(n-3)$ cross-ratios $(u_{2,n},...,u_{n-2,n})$.
We have
\bq
 d^{n-3}z & = & \left( \prod\limits_{j=3}^{n-2} u_{j,n}^{j-2} \right) d^{n-3}u.
\eq
Let us now fix $i_0 \in \{2,...,n-2\}$. We will study the limit $u_{i_0,n} \rightarrow 0$.
The chord $(i_0,n)$ splits the polygon into two smaller polygons. 
We set $z'=(z_1,z_2,...,z_{i_0})$ and $z''=(z_{i_0+1},...,z_n)$.
As before we label the new edge by $z_e$.
One of the two smaller polygons has the edges $z' \cup \{z_e\}$ 
and the dihedral structure $\pi'=(1,2,...,i_0,e)$,
the other smaller polygon has the edges $z'' \cup \{z_e\}$ and
the dihedral structure $\pi''=(e,i_0+1,i_0+2,...,n)$.
In the limit $u_{i_0,n} \rightarrow 0$ we have
\bq
 \lim\limits_{u_{i_0,n} \rightarrow 0} u_{i,j} & = & 1
\eq
for any chord $(i,j) \in \chi(z,\pi)$ which crosses the chord $(i_0,n) \in \chi(z,\pi)$.

Let us express the cyclic factor $C(\pi,z)$ in the new variables $u_{j,n}$.
We have
\bq
 \lim\limits_{z_n \rightarrow \infty} z_n^2 \; C\left(\pi,z\right)
 & = &
 \prod\limits_{j=2}^{n-2} u_{j,n}^{1-j} \left( u_{j,n} - 1 \right)^{-1}. 
\eq
Thus
\bq
 C\left(\pi,z\right)
 \frac{d^nz}{d\omega}
 & = &
 \left( \prod\limits_{j=2}^{n-2} \frac{1}{u_{j,n}\left( u_{j,n} - 1 \right)} \right) d^{n-3}u.
\eq

\section{The scattering forms}
\label{sect:scattering_forms}

In this section we study the scattering forms
\bq
 \Omega_{\mathrm{scattering}}^{\mathrm{cyclic}}\left(\sigma,z\right) 
 & = &
  C\left(\sigma,z\right) \frac{d^nz}{d\omega},
 \nonumber \\
 \Omega_{\mathrm{scattering}}^{\mathrm{pol}}\left(p,\eps,z\right)
 & = &
  E\left(p,\eps,z\right) \frac{d^nz}{d\omega}.
\eq
The polarisation factor $E(p,\eps,z)$ will be defined below.
These scattering forms have interesting properties:
\begin{enumerate}
\item The scattering forms are $\mathrm{PSL}(2,{\mathbb C})$-invariant.
\item The twisted intersection numbers give the amplitudes for the bi-adjoint scalar theory (cyclic, cyclic),
Yang-Mills theory (cyclic,polarisation) and gravity (polarisation,polarisation).
\item The only singularities of the scattering forms are on the divisor $\overline{\mathcal M}_{0,n} \backslash {\mathcal M}_{0,n}$.
\item The singularities are logarithmic.
\item The residues at the singularities factorise into two scattering forms of lower points.
\end{enumerate}
In addition, the scattering forms have specific properties under permutations:
The scattering form $\Omega_{\mathrm{scattering}}^{\mathrm{cyclic}}\left(\sigma,z\right)$
is invariant under cyclic permutations and satisfies the Kleiss-Kuijf relations \cite{Kleiss:1988ne}.
The scattering form $\Omega_{\mathrm{scattering}}^{\mathrm{pol}}\left(p,\eps,z\right)$ is permutation-invariant.

The most remarkable property is certainly the factorisation of the residues.
Please note that we do not require a particular kinematic limit in the momenta variables, like one momentum soft or
$s_I \rightarrow 0$ for a subset $I$ of $\{1,2,...,n\}$.
The factorisation of the residues of the scattering forms holds for any momentum configuration $p$.
\begin{figure}
\begin{center}
\includegraphics[scale=0.6]{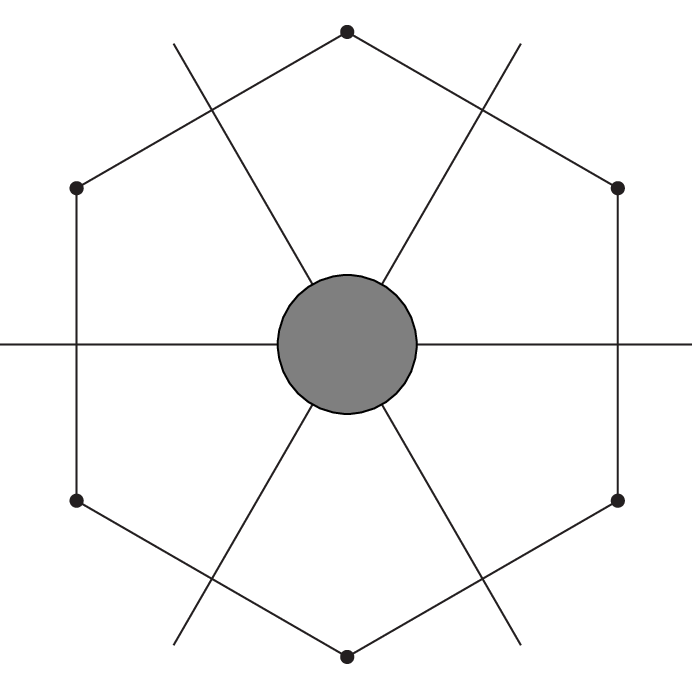}
\hspace*{20mm}
\includegraphics[scale=0.6]{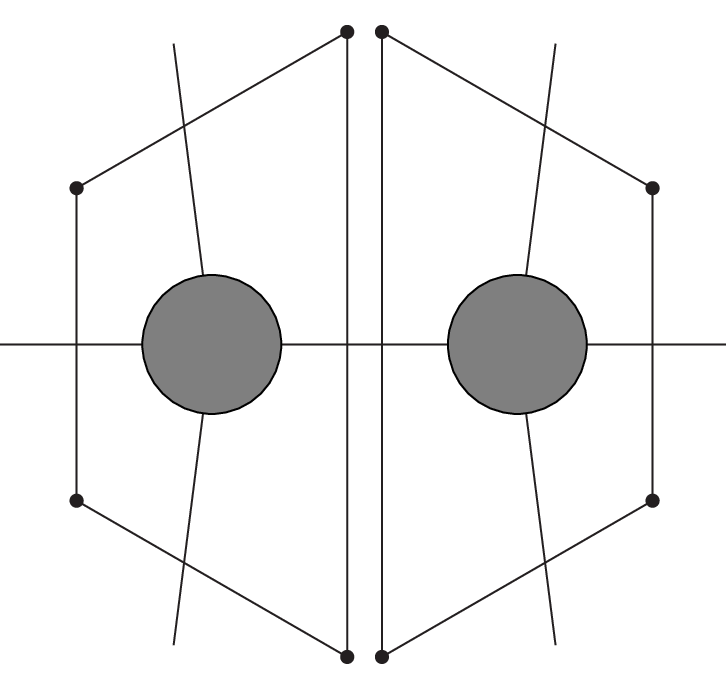}
\end{center}
\caption{
We may visualise a scattering form on $\mathcal M_{0,n}^\pi$ as shown in the left figure:
The  dihedral structure defines an $n$-gon, each external line of the scattering process
crosses one edge of the $n$-gon.
In the limit $u_{i_0,n}\rightarrow 0$ the residue factorises into two scattering forms of lower points, as shown in the
right figure.
}
\label{fig_factorisation}
\end{figure}
The factorisation of the residues is illustrated in fig.~(\ref{fig_factorisation}).
We may iterate this procedure and take multiple residues.
Taking the maximal number of residues produces a $0$-form and corresponds to a triangulation
of the $n$-gon.
\begin{figure}
\begin{center}
\includegraphics[scale=0.8]{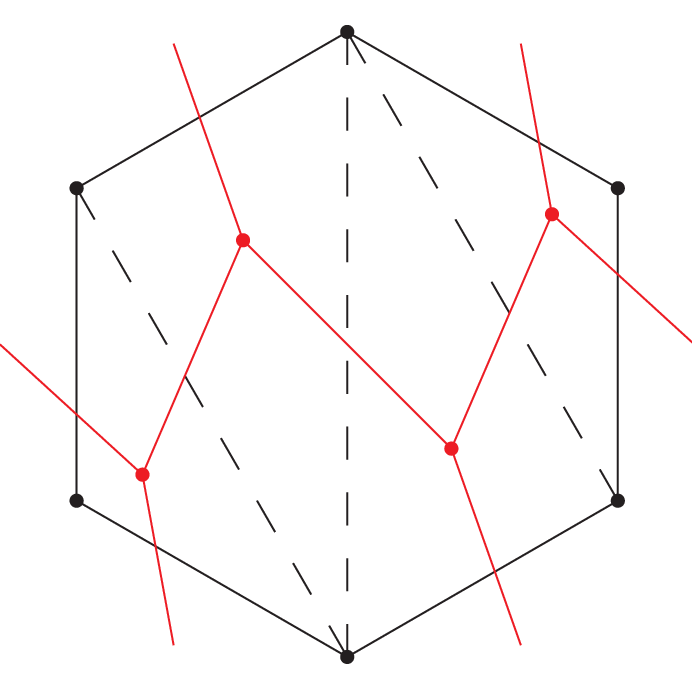}
\end{center}
\caption{
A hexagon with a triangulation, given by the dashed lines. The dual graph is shown in red.
}
\label{fig_triangulation}
\end{figure}
Each triangulation defines a dual graph $G$ with trivalent vertices, as shown in fig.~(\ref{fig_triangulation}),
and conversely, every dual graph with trivalent vertices defines a triangulation.
The $(n-3)$-fold maximal residue is given by the numerator of the dual graph.
This graph factorises into three-valent vertices,
where a sum over internal polarisations and particle flavours is understood.

Property $2$ has been shown recently by Mizera \cite{Mizera:2017rqa}. 
We believe that properties $3-5$ for $\Omega_{\mathrm{scattering}}^{\mathrm{cyclic}}$ 
are known to experts in the field, the fact that the singularities of $\Omega_{\mathrm{scattering}}^{\mathrm{cyclic}}$
are logarithmic is shown in \cite{Mizera:2017cqs}.
These properties can also be inferred from the conference talks \cite{Arkani-Hamed:amplitudes2017,Bai:amplitudes2017}.
The essential new ingredient of this paper is that $\Omega_{\mathrm{scattering}}^{\mathrm{pol}}$ can be defined with the same
properties.

\subsection{The cyclic scattering form}
\label{sect:cyclic_scattering_form}

Let us start with
\bq
 \Omega_{\mathrm{scattering}}^{\mathrm{cyclic}}\left(\sigma,z\right) 
 & = &
  C\left(\sigma,z\right) \frac{d^nz}{d\omega}.
\eq
$\mathrm{PSL}(2,{\mathbb C})$-invariance is straightforward.
Property $(2)$ is the statement that
with 
\bq
 \Omega \;\, = \;\;
 \Omega_{\mathrm{scattering}}^{\mathrm{cyclic}}\left(\sigma,z\right),
 & &
 \tilde{\Omega} \;\, = \;\;
 \Omega_{\mathrm{scattering}}^{\mathrm{cyclic}}\left(\tilde{\sigma},z\right)
\eq
one has
\bq
 m_n\left(\sigma,\tilde{\sigma},p\right)
 & = &
 i \oint\limits_{\mathcal C} d\Omega_{\mathrm{CHY}} \; C\left(\sigma,z\right) \; C\left(\tilde{\sigma},z\right)
 \;\; = \;\;
 i \left( \Omega, \tilde{\Omega} \right)_\eta,
\eq
where $(\Omega,\tilde{\Omega})_\eta$ denotes the twisted intersection number of $\Omega$ and $\tilde{\Omega}$
with twist $\eta$ (defined in eq.~(\ref{def_eta})).
The last equality has been shown recently by Mizera \cite{Mizera:2017rqa}.

Property $(3)$ is also clear: The cyclic factor $C(\sigma,z)$ becomes singular whenever two $z$'s adjacent in the cyclic order $\sigma$ coincide:
\bq
 z_{\sigma_i} & = & z_{\sigma_{i+1}}.
\eq
These points are on the divisor $\overline{\mathcal M}_{0,n} \backslash {\mathcal M}_{0,n}$.

In order to show points $(4)$ and $(5)$ we have to work a little bit more.
We consider the dihedral extension $\mathcal M_{0,n}^\pi$.
Without loss of generality we may take $\pi=(1,2,...,n)$.
Further let $i_0 \in \{2,...,n-2\}$.
The variable $i_0$ defines a chord $(i_0,n)$ and a cross-ratio $u_{i_0,n}$.
We now consider the behaviour of $\Omega_{\mathrm{scattering}}^{\mathrm{cyclic}}(\sigma,z)$
in the limit $u_{i_0,n} \rightarrow 0$.
Since we fixed $\pi=(1,2,...,n)$, the other permutation $\sigma=(\sigma_1,...,\sigma_n)$
is arbitrary.
Let us call 
\bq
 z_{\sigma_i} - z_{\sigma_j}
\eq
a bond connecting the edges $z_{\sigma_i}$ and $z_{\sigma_j}$.
\begin{figure}
\begin{center}
\includegraphics[scale=0.8]{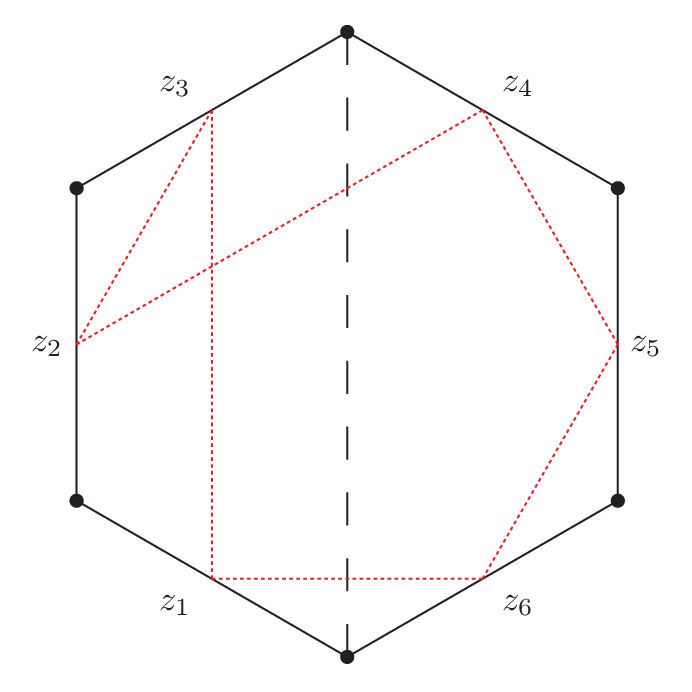}
\end{center}
\caption{
A hexagon with the dihedral structure $\pi=(1,2,3,4,5,6)$.
The dashed line shows the chord $(3,6)$.
The dotted red lines show the bonds corresponding to the cyclic factor $C(\sigma,z)$ for
$\sigma=(1,3,2,4,5,6)$.
The bonds $(z_2-z_4)$ and $(z_6-z_1)$ cross the chord $(3,6)$.
}
\label{fig_bonds}
\end{figure}
This is illustrated in fig.~\ref{fig_bonds}.
We say that a bond $(z_{\sigma_i}-z_{\sigma_j})$ crosses the chord $(i_0,n)$ if
\bq
 \left( \sigma_i \in \{1,...,i_0\} \; \mathrm{and} \; \sigma_j \in \{i_0+1,...,n\} \right)
 \;\;\; \mathrm{or} \;\;\;
 \left( \sigma_j \in \{1,...,i_0\} \; \mathrm{and} \; \sigma_i \in \{i_0+1,...,n\} \right).
 \nonumber
\eq
Now let us look at the cyclic factor
\bq
 C\left(\sigma,z\right)
 & = &
 \frac{1}{\left(z_{\sigma_1}-z_{\sigma_2}\right) ... \left(z_{\sigma_{n-1}}-z_{\sigma_n}\right) \left(z_{\sigma_n}-z_{\sigma_1}\right)}.
\eq
We are interested in the number of bonds crossing the chord $(i_0,n)$.
It is easy to see that this number must be even and that there are at least two bonds
crossing the chord $(i_0,n)$.
The maximal number of bonds crossing the chord $(i_0,n)$  is given by
\bq
 2 \min\left(i_0,n-i_0\right).
\eq
We call $\sigma$ and $\pi$ equivalent with respect to the chord $(i_0,n)$ 
\bq
 \sigma & \left.\sim\right._{(i_0,n)} & \pi,
\eq
if there are exactly
two bonds in $C(\sigma,z)$ crossing the chord $(i_0,n)$.
It is not too difficult to see that this is the case if and only if $\sigma$ can be written
(after a suitable cyclic permutation) in the form
\bq
 \sigma & = & \left( \sigma_1', ..., \sigma_{i_0}', \sigma_{i_0+1}'', ..., \sigma_n'' \right)
\eq
with
\bq
 \sigma_j' \in \{1,...,i_0\}
 & \mbox{and} &
 \sigma_k'' \in \{i_0+1,...,n\}.
\eq
If $\sigma$ and $\pi$ are equivalent with respect to the chord $(i_0,n)$ we define the induced
dihedral structures $\sigma'$ and $\sigma''$ to be
\bq
 \sigma' \; = \; \left( \sigma_1', ..., \sigma_{i_0}', e \right),
 & &
 \sigma'' \; = \; \left( e, \sigma_{i_0+1}'', ..., \sigma_n'' \right).
\eq
These considerations are helpful to answer the following question: 
How many factors of $u_{i_0,n}$ does the cyclic factor $C(\sigma,z)$ produce, if we change the
variables from the $z_j$'s to the cross-ratios $u_{j,n}$'s?
A bond 
\bq
 z_{\sigma_i} - z_{\sigma_j}
\eq
gives a factor $u_{i_0,n}$ if $\sigma_i, \sigma_j \in \{1,...,i_0\}$ and no factor
of $u_{i_0,n}$ in all other cases.
Thus $C(\sigma,z)$ has a factor
\bq
 u_{i_0,n}^{\frac{n_{\mathrm{cross}}}{2}-i_0},
\eq
where $n_{\mathrm{cross}}$ denotes the number of bonds crossing the chord $(i_0,n)$.
We obtain the maximal (negative) power of $u_{i_0,n}$ 
\bq
 u_{i_0,n}^{1-i_0},
\eq
if $\sigma$ is equivalent to $\pi$ with respect
to the chord $(i_0,n)$.
In all other cases we obtain fewer powers of $1/u_{i_0,n}$.
Combined with the factor
\bq
 u_{i_0,n}^{i_0-2}
\eq
from the measure we obtain a single pole $1/u_{i_0,n}$ if 
$\sigma$ is equivalent to $\pi$ with respect
to the chord $(i_0,n)$, and no pole in all other cases.
This proves property $(4)$.

Let us now look at the residues.
If $\sigma \nsim_{(i_0,n)} \pi$ there aren't enough negative powers of $u_{i_0,n}$ from the cyclic factors $C(\sigma,z)$
to produce a non-zero residue.
If $\sigma \sim_{(i_0,n)} \pi$ 
the scattering form $\Omega_{\mathrm{scattering}}^{\mathrm{cyclic}}(\sigma,z)$
has a single pole at $u_{i_0,n}=0$.
The chord $(i_0,n)$ divides the polygon with dihedral structure $\pi$ into two smaller polygons
with dihedral structures $\pi'$ and $\pi''$, respectively.
As $\sigma$ is equivalent to $\pi$ with respect to the chord $(i_0,n)$
we have the induced structures $\sigma'$ and $\sigma''$.
On the polygon with dihedral structure $\pi'$ we use the coordinates
$u_{2,n}, ..., u_{i_0-1,n}$,
on the polygon with dihedral structure $\pi''$ we use the coordinates
$u_{i_0+1,n}, ..., u_{n-2,n}$.
We denote by $Y$ be the hypersurface given by $u_{i_0,n}=0$.
After cancelling common factors of $(u_{i_0+1,n} u_{i_0+2,n} ... u_{n-2,n})$
from the numerator and the denominator one finds
\bq
 \mathrm{Res}_Y \Omega_{\mathrm{scattering}}^{\mathrm{cyclic}}\left(\sigma,z\right)
 & = &
 \left\{
 \begin{array}{ll}
  \left(-1\right)^{i_0-1}
  \Omega_{\mathrm{scattering}}^{\mathrm{cyclic}}\left(\sigma',z\right)
   \wedge
  \Omega_{\mathrm{scattering}}^{\mathrm{cyclic}}\left(\sigma'',z\right)
 & \mbox{if} \;\; \sigma \sim_{(i_0,n)} \pi, \\
 0 & \mbox{otherwise}. \\
 \end{array}
 \right. \nonumber 
 \\
\eq
Thus we see that the residue factorises into two scattering forms of lower points.
The factorisation of the residue is illustrated in fig.~(\ref{fig_factorisation}).

The scattering form $\Omega_{\mathrm{scattering}}^{\mathrm{cyclic}}(\sigma,z)$
inherits the transformation properties under permutations from the cyclic factor $C(\sigma,z)$.
It is invariant under cyclic permutations and satisfies the 
Kleiss-Kuijf relations \cite{Kleiss:1988ne}.
Let $\alpha=(\alpha_1,...,\alpha_j)$ be a permutation of $(2,...,j+1)$ and
$\beta=(\beta_1,...,\beta_{n-2-j})$ a permutation of $(j+2,...,n-1)$.
Then
\bq
\label{Kleiss_Kuijf_relation}
 \Omega_{\mathrm{scattering}}^{\mathrm{cyclic}}\left((1,\alpha,n,\beta),z\right)
 & = &
 \left(-1\right)^{n-2-j}
 \sum\limits_{\sigma \in \alpha \; \shuffle \; \beta^T}
 \Omega_{\mathrm{scattering}}^{\mathrm{cyclic}}\left((1,\sigma,n),z\right),
\eq
where $\alpha \; \shuffle \; \beta^T$ denotes the set of all shuffles of $\alpha$ with $\beta^T=(\beta_{n-2-j},...,\beta_1)$, i.e.
the set of all permutations of the elements of $\alpha$ and $\beta^T$, which preserve the relative order of the
elements of $\alpha$ and of the elements of $\beta^T$.

\subsubsection{Examples}

Let us give a few examples. We consider $\mathcal M_{0,n}^\pi$ with $\pi=(1,2,...,n)$.
We fix $z_1=0$, $z_{n-1}=1$ and $z_n=\infty$.
We have
\bq
 \frac{d^nz}{d\omega}
 & = &
 z_n \left( z_n-1 \right)
 dz_2 \wedge dz_3 \wedge ... \wedge dz_{n-2}.
\eq
Let us first discuss the case $n=3$. For three particles the scattering form
$\Omega_{\mathrm{scattering}}^{\mathrm{cyclic}}(\sigma,z)$ is a $0$-form.
On $\mathcal M_{0,3}^\pi$ the scattering form is given by
\bq
 \Omega_{\mathrm{scattering}}^{\mathrm{cyclic}}\left(\sigma,z\right) 
 & = &
 \pm 1,
\eq
depending on whether $\sigma$ is an even or odd permutation of $\pi=(1,2,3)$.

For $n=4$ and 
$\sigma=\pi=(1,2,3,4)$ we obtain
\bq
 \Omega_{\mathrm{scattering}}^{\mathrm{cyclic}}\left(\sigma,z\right)
 & = &
 \frac{dz_2}{z_2\left(z_2-1\right)}
 \;\; = \;\;
 \frac{du_{2,4}}{u_{2,4} \left(u_{2,4}-1\right)}.
\eq
The residue at $u_{2,4}=0$ is given by
\bq
 \mathrm{Res}_{u_{2,4}=0} \; \Omega_{\mathrm{scattering}}^{\mathrm{cyclic}}
 & = & -1.
\eq
Let us now consider the case $n=5$.
For $\sigma=\pi=(1,2,3,4,5)$ we obtain
\bq
 \Omega_{\mathrm{scattering}}^{\mathrm{cyclic}}\left(\sigma,z\right)
 & = &
 \frac{dz_2 \wedge dz_3}{z_2 \left( z_2-z_3 \right) \left(z_3-1\right)}
 \;\; = \;\;
 \frac{du_{2,5}}{u_{2,5} \left(u_{2,5}-1\right)}
 \wedge
 \frac{du_{3,5}}{u_{3,5} \left(u_{3,5}-1\right)}.
\eq
We have the residues
\bq
 \mathrm{Res}_{u_{2,5}=0} \; \Omega_{\mathrm{scattering}}^{\mathrm{cyclic}}
 \;\; = \;\;
 - \frac{du_{3,5}}{u_{3,5} \left(u_{3,5}-1\right)},
 & &
 \mathrm{Res}_{u_{3,5}=0} \; \Omega_{\mathrm{scattering}}^{\mathrm{cyclic}}
 \;\; = \;\;
 \frac{du_{2,5}}{u_{2,5} \left(u_{2,5}-1\right)}.
 \;\;
\eq
For general $n$ we have 
for $\sigma=\pi=(1,...,n)$
\bq
 \Omega_{\mathrm{scattering}}^{\mathrm{cyclic}}\left(\sigma,z\right)
 & = &
 \frac{dz_2 \wedge ... \wedge dz_{n-2}}{z_2 \left( z_2-z_3 \right) ... \left( z_{n-3}-z_{n-2} \right) \left(z_{n-2}-1\right)}
 \nonumber \\
 & = &
 \frac{du_{2,n}}{u_{2,n} \left(u_{2,n}-1\right)}
 \wedge ... \wedge
 \frac{du_{n-2,n}}{u_{n-2,n} \left(u_{n-2,n}-1\right)}.
\eq
For the residues we find
\bq
 \mathrm{Res}_{u_{i_0,n}=0} \; \Omega_{\mathrm{scattering}}^{\mathrm{cyclic}}\left(\sigma,z\right)
 & = &
 \left(-1\right)^{i_0-1}
 \frac{du_{2,n}}{u_{2,n} \left(u_{2,n}-1\right)}
 \wedge ... \wedge
 \frac{du_{i_0-1,n}}{u_{i_0-1,n} \left(u_{i_0-1,n}-1\right)}
 \nonumber \\
 & &
 \wedge
 \frac{du_{i_0+1,n}}{u_{i_0+1,n} \left(u_{i_0+1,n}-1\right)}
 \wedge ... \wedge
 \frac{du_{n-2,n}}{u_{n-2,n} \left(u_{n-2,n}-1\right)}.
\eq

\subsection{The polarisation scattering form}
\label{sect:polarisation_scattering_form}

In this paragraph we define and study the scattering form
\bq
 \Omega_{\mathrm{scattering}}^{\mathrm{pol}}\left(p,\eps,z\right)
 & = &
  E\left(p,\eps,z\right) \frac{d^nz}{d\omega}.
\eq
This scattering form involves the polarisation factor $E(p,\eps,z)$.
Originally, the polarisation factor $E(p,\eps,z)$ is defined in terms of the reduced Pfaffian
\bq
 \frac{\left(-1\right)^{i+j}}{2 \left(z_i-z_j\right)} \mathrm{Pf} \; \Psi^{ij}_{ij},
\eq
where $\Psi^{ij}_{ij}$ is the $(2n-2) \times (2n-2)$-matrix obtained from the $(2n)\times(2n)$-matrix
$\Psi$ by deleting the rows and columns $i$ and $j$ (with $1 \le i < j \le n$).
It can be shown that this definition is independent of the choice of $i$ and $j$ if
\begin{enumerate}
\item the momenta are on-shell: $p_j^2=0$,
\item the polarisations are transverse: $\eps_j \cdot p_j = 0$,
\item the variables $z$ are solutions of the scattering equations: $f_j(z,p)=0$.
\end{enumerate}
We will need to relax all three conditions.
The requirement to relax condition $(3)$ is obvious:
We want to study the polarisation factor $E(p,\eps,z)$ on the full moduli space
$\overline{\mathcal M}_{0,n}$ and not just on a zero-dimensional sub-variety defined by the solutions
of the scattering equations.
The requirements to relax conditions $(1)$ and $(2)$ become apparent once
we start to discuss the structure of possible factorisations.
Let us choose a dihedral structure, which we take without loss of generality as
$\pi=(1,2,...,n)$ and pick $i_0\in\{2,...,n-2\}$.
This defines a chord $(i_0,n)$, which
divides the original $n$-gon into two smaller polygons.
We set
\bq
\label{def_induced_momenta}
 p' & = & \left( p_1, p_2, ..., p_{i_0}, p_{i_0+1} + p_{i_0+2} + ... + p_n \right),
 \nonumber \\
 p'' & = & \left( p_1 + p_2 + ... + p_{i_0}, p_{i_0+1}, p_{i_0+2}, ..., p_n \right).
\eq
Note that the momenta of $p'$ and $p''$ are no longer necessarily on the mass-shell.
However, they satisfy momentum conservation.
Thus in general we have
\bq
 p_j^2 & \neq & 0.
\eq
For the polarisation vectors we set
\bq
\label{def_induced_polarisations}
 \eps' & = & \left( \eps_1, \eps_2, ..., \eps_{i_0}, \eps_e \right),
 \nonumber \\
 \eps'' & = & \left( \eps_e^\ast, \eps_{i_0+1}, \eps_{i_0+2}, ..., \eps_n \right),
\eq
where we introduced for the new edge defined by the chord $(i_0,n)$ a new polarisation vector $\eps_e$.
In four space-time dimensions massless gauge bosons have two physical polarisations, which we may take
as
\bq
\eps_{\mu}^{+}(p,q) = - \frac{\langle p+|\gamma_{\mu}|q+\rangle}{\sqrt{2} \langle p- | q + \rangle},
 & &
\eps_{\mu}^{-}(p,q) = \frac{\langle p-|\gamma_{\mu}|q-\rangle}{\sqrt{2} \langle p + | q - \rangle},
\eq
where $p$ is the momentum of the gauge boson and $q$ is an arbitrary light-like reference momentum.
The polarisation sum over the physical polarisation is given by
\bq
\sum\limits_{\lambda \in \{+,-\} } \left( \eps_\mu^\lambda \right)^\ast \; \eps_\nu^\lambda
 & = & 
 - g_{\mu\nu} + \frac{p_\mu q_\nu}{p \cdot q} + \frac{q_\mu p_\nu}{p \cdot q}.
\eq 
On both sides we have a $4 \times 4$-matrix (in $\mu$, $\nu$) of rank $2$.
In order to obtain a matrix of full rank, let us supplement the two physical polarisations $\eps^+$ and $\eps^-$
by two un-physical polarisations $\eps^0$ and $\eps^{\bar{0}}$, sucht that
\bq
\sum\limits_{\lambda \in \{+,-,0,\bar{0}\} } \left( \eps_\mu^\lambda \right)^\ast \; \eps_\nu^\lambda
 & = & 
 - g_{\mu\nu}.
\eq 
In $D$ space-time dimensions we have $(D-2)$ physical polarisations and $2$ un-physical polarisations.
In this paper we will never need to specify the explicit expressions for the polarisation vectors.
We only need to assume that in arbitrary space-time dimensions there is a set of (physical and un-physical)
polarisation vectors indexed by $\lambda$ such that
\bq
\sum\limits_{\lambda} \left( \eps_\mu^\lambda \right)^\ast \; \eps_\nu^\lambda
 & = & 
 - g_{\mu\nu}.
\eq 
With this generalisation we now have in general
\bq
 \eps_j \cdot p_j & \neq & 0.
\eq
Analogously we may introduce (un-physical) polarisations $\eps^\lambda_{\mu\nu}$ for the auxiliary tensor particle appearing in eq.~(\ref{tensor_particle_propagator}) and eq.~(\ref{tensor_particle_vertex}),
such that
\bq
\sum\limits_{\lambda} \left( \eps_{\mu\nu}^\lambda \right)^\ast \; \eps_{\rho\sigma}^\lambda
 & = &
 - \frac{1}{2} p^2 \left( g_{\mu\rho} g_{\nu\sigma} - g_{\mu\sigma} g_{\nu\rho} \right).
\eq
We may iterate this procedure for all new vertices appearing in the effective Lagrangian
of eq.~(\ref{YM_vers2}).
As an example we discuss a five-valent vertex, appearing in ${\mathcal L}^{(5)}$.
\begin{figure}
\begin{center}
\includegraphics[scale=0.7]{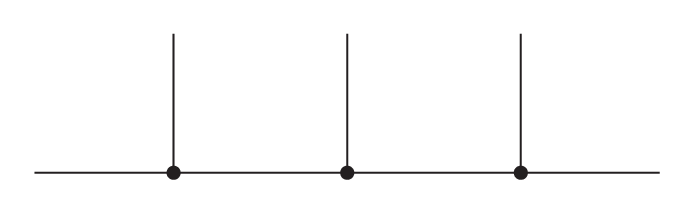}
\includegraphics[scale=0.7]{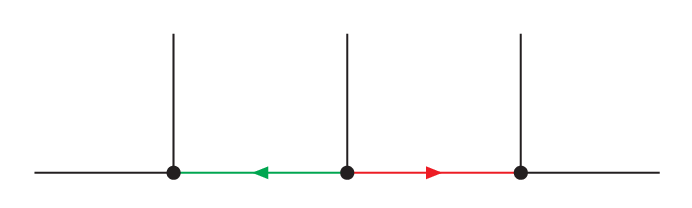}
\includegraphics[scale=0.7]{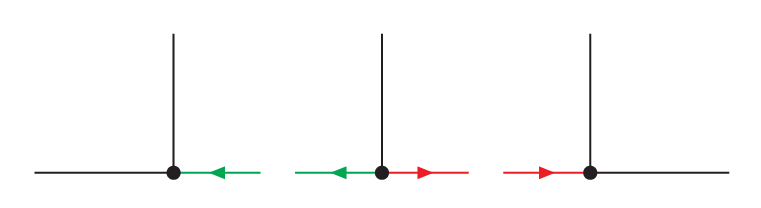}
\end{center}
\caption{
The left picture shows the tree structure underlying a five-valent vertex from ${\mathcal L}^{(5)}$.
For the intermediate propagators we may introduce auxiliary particles, shown in green and red
in the middle picture.
Introducing quantum numbers $g$, $\bar{g}$ and $r$, $\bar{r}$ such that the new auxiliary particles propagate only from $g$ to $\bar{g}$ and from $r$ to $\bar{r}$ ensures that the new 
three-valent vertices (shown in the right picture)
recombine only into the original five-valent vertex.
}
\label{fig_five_vertex}
\end{figure}
This vertex has an underlying tree structure as shown in the left picture 
of fig.~(\ref{fig_five_vertex}).
(For five points there is only one possible underlying tree structure.)
We may introduce two auxiliary particles for the intermediate edges, as shown in the middle picture
of fig.~(\ref{fig_five_vertex}).
We may associate quantum numbers to these particles, say $g$ and $r$, such that the propagation
is only from $g$ to $\bar{g}$ and from $r$ to $\bar{r}$.
In this way we obtain three new three-valent vertices, as shown in the right picture
of fig.~(\ref{fig_five_vertex}).
Note that the introduction of the quantum numbers $g$ and $r$ ensures, that the new auxiliary vertices
and propagators recombine only into the original five-valent vertex.

In the following we will write
\bq
 \sum\limits_{f, \; \lambda}
\eq
for a sum over particle species $f$ (gauge boson and auxiliary particles) and the corresponding polarisations $\lambda$.
By a factorisation of a numerator $N(G)$ corresponding to a graph $G=(G_1,G_2)$ we understand
a factorisation of the form
\bq
 N\left(G\right)
 & = &
 \sum\limits_{f, \; \lambda} N\left(G_1\right) N\left(G_2\right),
\eq
where the sum is over all particles and polarisations corresponding to the edge connecting the sub-graphs
$G_1$ and $G_2$.

The reduced Pfaffian has been studied quite extensively 
in the literature \cite{Du:2013sha,Litsey:2013jfa,Lam:2016tlk,Bjerrum-Bohr:2016axv,Huang:2017ydz,Du:2017kpo,Gao:2017dek,Chen:2016fgi,Chen:2017edo,Chen:2017bug}.
The expression given by the reduced Pfaffian is not very well suited to generalise towards 
off-shell momenta, 
unphysical polarisation or away from the solutions of the scattering equations.
Appendix~\ref{counter_example} provides further details on this point.
In order to construct the polarisation factor $E(p,\eps,z)$ in the general off-shell case with
unphysical polarisations and away from the solutions of the scattering equations we proceed along 
a different way.
As polarisation factor we may take
\bq
\label{def_polarisation_factor_1}
 E\left(p,\eps,z\right)
 & = &
 \sum\limits_{\kappa \in S_{n-2}^{(i,j)}}
 C\left(\kappa,z\right)
 \;
 N^{\mathrm{BCJ}}_{\mathrm{comb}}\left(\kappa\right),
\eq
where $i,j \in \{1,...,n\}$, $i \neq j$ and $\kappa$ is a permutation of $\{1,...,n\}$ with $\kappa_1=i$
and $\kappa_n=j$.
The set of all these permutations is denoted by $S_{n-2}^{(i,j)}$.
The BCJ-numerators $N^{\mathrm{BCJ}}_{\mathrm{comb}}(\kappa)$ have been defined in eq.~(\ref{def_BCJ_numerator}).
Eq.~(\ref{def_polarisation_factor_1}) is clearly invariant under the $(n-2)!$ 
permutations of $\{1,...,n\} \backslash \{i,j\}$.
It can be shown that eq.~(\ref{def_polarisation_factor_1}) is actually invariant under all $n!$ permutations of
$\{1,...,n\}$ and therefore independent of the choice of $i$ and $j$.
A proof is given in appendix~\ref{sect:perm_invariance}.
Our standard choice will be $i=1$ and $j=n$, yielding
\bq
\label{def_Omega_pol}
 \Omega_{\mathrm{scattering}}^{\mathrm{pol}}\left(p,\eps,z\right)
 & = &
 E\left(p,\eps,z\right)
 \frac{d^nz}{d\omega},
 \nonumber \\
 E\left(p,\eps,z\right)
 & = &
 \sum\limits_{\kappa \in S_{n-2}^{(1,n)}}
 C\left(\kappa,z\right)
 \;
 N^{\mathrm{BCJ}}_{\mathrm{comb}}\left(\kappa\right),
\eq
where the sum is now over all $(n-2)!$ permutations of $\{2,...,(n-1)\}$, keeping $\kappa_1=1$ and
$\kappa_n=n$ fixed.

Let us now discuss the properties of $\Omega_{\mathrm{scattering}}^{\mathrm{pol}}$.
As already mentioned, the polarisation factor $E(p,\eps,z)$ is permutation-invariant, hence 
$\Omega_{\mathrm{scattering}}^{\mathrm{pol}}$ is permutation-invariant.

Eq.~(\ref{def_Omega_pol}) is an expansion in cyclic factors.
This has three immediate implications.
First of all, $\mathrm{PSL}(2,{\mathbb C})$-invariance is manifest.
Secondly, it follows that the only singularities of $\Omega_{\mathrm{scattering}}^{\mathrm{pol}}$ 
occur when two of the $z$'s coincide, thus
on the divisor $\overline{\mathcal M}_{0,n} \backslash {\mathcal M}_{0,n}$.
Thirdly, from the discussion in section~\ref{sect:cyclic_scattering_form} it follows
that all singularities are logarithmic.

It is a little bit harder to show the remaining two properties:
Firstly, we have to show that $E(p,\eps,z)$ as defined by eq.~(\ref{def_Omega_pol})
agrees with the reduced Pfaffian in the on-shell limit with physical polarisations and on the sub-variety defined by the scattering equations.
Secondly, we have to establish the factorisation of the residues.

Let us start with the equivalence of $E(p,\eps,z)$ with the reduced Pfaffian
for on-shell momenta, physical polarisations and on the sub-variety defined by the scattering equations.
Let us therefore assume on-shell momenta and physical polarisations.
$E(p,\eps,z)$ agrees with the reduced Pfaffian for all $(n-3)!$ solutions of the scattering equations $z^{(j)}$
if and only if
\bq
\label{verification_A_n}
 i \oint\limits_{\mathcal C} d\Omega_{\mathrm{CHY}} \; C\left(\sigma,z\right) E\left(p,\eps,z\right)
\eq
reproduces the Yang-Mills amplitude $A_n(\sigma,p,\eps)$ for a basis of $(n-3)!$ cyclic orders $\sigma$.
Since $E(p,\eps,z)$ is permutation-invariant it suffices to check eq.~(\ref{verification_A_n}) for one cyclic order $\sigma$.
It is convenient to take $\sigma=(1,2,...,n)$.
From eq.~(\ref{def_N_BCJ}) we have
\bq
\label{A_n_lhs}
 A_n\left(\sigma,p,\eps\right)
 & = &
 i \left(-1\right)^{n-3}
 \sum\limits_{G \in {\mathcal T}_n(\sigma)} N^{\mathrm{BCJ}}(G) \prod\limits_{e \in E(G)} \frac{1}{s_e}.
\eq
Working out eq.~(\ref{verification_A_n}) we obtain
\bq
\label{A_n_rhs}
\lefteqn{
 i \oint\limits_{\mathcal C} d\Omega_{\mathrm{CHY}} \; C\left(\sigma,z\right) E\left(p,\eps,z\right)
 = 
 } & & \\
 & & 
 i \left(-1\right)^{n-3}
 \sum\limits_{G \in {\mathcal T}_n(\sigma)} 
 \left( \prod\limits_{e \in E(G)} \frac{1}{s_e} \right)
 \sum\limits_{\tilde{\sigma} \in \mathrm{CO}(G)}
 \sum\limits_{\kappa \in S_{n-2}^{(1,n)}}
 \delta_{\tilde{\sigma},\kappa}
 \left(-1\right)^{n_{\mathrm{flip}}(\sigma,\kappa)}
 N^{\mathrm{BCJ}}_{\mathrm{comb}}\left(\kappa\right),
 \nonumber
\eq
where we used a formula similar to eq.~(\ref{exchange_summations}) to exchange summation orders
and where we used eq.~(\ref{def_m_n}).
The symbol $\delta_{\sigma,\sigma'}$ equals 
$1$ if $\sigma=\sigma'$ as $n$-tuples and zero otherwise.
Let us denote
\bq
 \hat{N}^{\mathrm{BCJ}}(G)
 & = &
 \sum\limits_{\tilde{\sigma} \in \mathrm{CO}(G)}
 \sum\limits_{\kappa \in S_{n-2}^{(1,n)}}
 \delta_{\tilde{\sigma},\kappa}
 \left(-1\right)^{n_{\mathrm{flip}}(\sigma,\kappa)}
 N^{\mathrm{BCJ}}_{\mathrm{comb}}\left(\kappa\right).
\eq
Eq.~(\ref{A_n_lhs}) and eq.~(\ref{A_n_rhs}) are certainly equal if
\bq
\label{numerator_equality}
 \hat{N}^{\mathrm{BCJ}}(G) & = & N^{\mathrm{BCJ}}(G).
\eq
Actually, it would be sufficient if $\hat{N}^{\mathrm{BCJ}}(G)$ and $N^{\mathrm{BCJ}}(G)$
are related by a generalised gauge transformation, i.e.
\bq
 \sum\limits_{G \in {\mathcal T}_n(\sigma)} 
 \frac{\hat{N}^{\mathrm{BCJ}}(G)-N^{\mathrm{BCJ}}(G)}{\prod\limits_{e \in E(G)} s_e}
 & = & 0.
\eq
However, for the case at hand one can show that the stronger condition of eq.~(\ref{numerator_equality}) holds.
In fact, $\hat{N}^{\mathrm{BCJ}}(G)$ is nothing else than the reduction of the numerator $N^{\mathrm{BCJ}}(G)$
to the basis of BCJ-numerators $N^{\mathrm{BCJ}}_{\mathrm{comb}}\left(\kappa\right)$
with $\kappa \in S_{n-2}^{(1,n)}$ by repeated use of eq.~(\ref{STU_relation}).
The details are given in appendix~\ref{sect:BCJ_numerator_reduction}.
Having established that the definition of the polarisation factor in eq.~(\ref{def_polarisation_factor_1}) agrees
with the reduced Pfaffian
for on-shell momenta, physical polarisations and on the sub-variety defined by the
scattering equations, we may use the expression of eq.~(\ref{def_polarisation_factor_1}) also for the gravity
amplitudes.

Let us now look at the factorisation of the residues.
We consider the dihedral extension $\mathcal M_{0,n}^\pi$.
Without loss of generality we may take $\pi=(1,2,...,n)$.
Further let $i_0 \in \{2,...,n-2\}$.
The variable $i_0$ defines a chord $(i_0,n)$ and a cross-ratio $u_{i_0,n}$.
We now consider the behaviour of $\Omega_{\mathrm{scattering}}^{\mathrm{pol}}(p,\eps,z)$
in the limit $u_{i_0,n} \rightarrow 0$.
Let us denote by $Y$ the hypersurface given by $u_{i_0,n}=0$.
The chord $(i_0,n)$ divides the $n$-gon defined by $\pi$ into two smaller polygons.
We denote by $p'$ and $\eps'$ the induced data of one of the two smaller polygons,
and by $p''$ and $\eps''$ the induced data of the other smaller polygon.
The new momenta $p'$ and $p''$ are defined by eq.~(\ref{def_induced_momenta}),
the new polarisations $\eps'$ and $\eps''$ according to eq.~(\ref{def_induced_polarisations}).
Let us introduce the set of indices
\bq
 I' \;\; = \;\; \left\{ 1, 2, ..., i_0, e \right\},
 & &
 I'' \;\; = \;\; \left\{ e, i_0+1, i_0+2, ..., n \right\}.
\eq
We denote by $S_{i_0-1}^{(1,e)}$ 
the set of permutations $\kappa'$ of the set $I'$, where the values $\kappa_1'=1$ and $\kappa_{i_0+1}'=e$
are fixed.
Similarly, we denote by $S_{n-i_0-1}^{(e,n)}$
the set of permutations $\kappa''$ of the set $I''$, where the values $\kappa_1''=e$ and $\kappa_{n-i_0+1}''=n$ are fixed.
For 
\bq
 \kappa' \; = \; \left( 1, \kappa_2', ..., \kappa_{i_0}', e \right) \in S_{i_0-1}^{(1,e)}, 
 & &
 \kappa'' \; = \; \left( e, \kappa_{i_0+1}'', ..., \kappa_{n-1}'', n \right) \in S_{n-i_0-1}^{(e,n)}
\eq
we define $\kappa = (\kappa',\kappa'')$ by
\bq
 \kappa & = &
 \left( 1, \kappa_2', ..., \kappa_{i_0}', \kappa_{i_0+1}'', ..., \kappa_{n-1}'', n \right).
\eq
We would like to show
\bq
\label{Omega_E_factorisation}
 \mathrm{Res}_Y \Omega_{\mathrm{scattering}}^{\mathrm{pol}}\left(p,\eps,z\right)
 & = &
  \sum\limits_{f, \; \lambda} 
  \left(-1\right)^{i_0-1}
  \Omega_{\mathrm{scattering}}^{\mathrm{pol}}\left(p',\eps',z\right)
   \wedge
  \Omega_{\mathrm{scattering}}^{\mathrm{pol}}\left(p'',\eps'',z\right).
\eq
Let us first look at the left-hand side.
From the definition in eq.~(\ref{def_Omega_pol}) we have
\bq
 \mathrm{Res}_Y \Omega_{\mathrm{scattering}}^{\mathrm{pol}}\left(p,\eps,z\right)
 & = &
 \sum\limits_{\kappa \in S_{n-2}^{(1,n)}}
 N^{\mathrm{BCJ}}_{\mathrm{comb}}\left(\kappa\right)
 \mathrm{Res}_Y \Omega_{\mathrm{cyclic}}^{\mathrm{pol}}\left(\kappa,z\right).
\eq
From section~\ref{sect:cyclic_scattering_form} we know that there is only a residue
if $\kappa \sim_{(i_0,n)} \pi$. This is the case, if $\kappa$ can be written as
$\kappa=(\kappa',\kappa'')$ with $\kappa' \in S_{i_0-1}^{(1,e)}$
and $\kappa'' \in S_{n-i_0-1}^{(e,n)}$.
Therefore
\bq
\label{eq_lhs}
 \mbox{l.h.s.}
 & = &
 \left(-1\right)^{i_0-1}
 \sum\limits_{\kappa' \in S_{i_0-1}^{(1,e)}}
 \sum\limits_{\kappa'' \in S_{n-i_0-1}^{(e,n)}}
 N^{\mathrm{BCJ}}_{\mathrm{comb}}\left((\kappa',\kappa'')\right)
  \Omega_{\mathrm{scattering}}^{\mathrm{cyclic}}\left(\kappa',z\right)
   \wedge
  \Omega_{\mathrm{scattering}}^{\mathrm{cyclic}}\left(\kappa'',z\right).
 \nonumber \\
\eq
Now, $N^{\mathrm{BCJ}}_{\mathrm{comb}}((\kappa',\kappa''))$
is the BCJ-numerator of a multi-peripheral graph and factorises as
\bq
 N^{\mathrm{BCJ}}_{\mathrm{comb}}\left((\kappa',\kappa'')\right)
 & = &
  \sum\limits_{f, \; \lambda} 
 N^{\mathrm{BCJ}}_{\mathrm{comb}}\left(\kappa'\right)
 N^{\mathrm{BCJ}}_{\mathrm{comb}}\left(\kappa''\right).
\eq
Therefore
\bq
\lefteqn{
 \mbox{l.h.s.}
 = 
 } & & \nonumber \\
 & &
 \left(-1\right)^{i_0-1}
 \sum\limits_{f, \; \lambda} 
 \sum\limits_{\kappa' \in S_{i_0-1}^{(1,e)}}
 N^{\mathrm{BCJ}}_{\mathrm{comb}}\left(\kappa'\right)
 \Omega_{\mathrm{scattering}}^{\mathrm{cyclic}}\left(\kappa',z\right)
 \wedge
 \sum\limits_{\kappa'' \in S_{n-i_0-1}^{(e,n)}}
 N^{\mathrm{BCJ}}_{\mathrm{comb}}\left(\kappa''\right)
 \Omega_{\mathrm{scattering}}^{\mathrm{cyclic}}\left(\kappa'',z\right),
 \nonumber
\eq
and hence
\bq
 \mbox{l.h.s.}
 & = &
 \sum\limits_{f, \; \lambda} 
 \left(-1\right)^{i_0-1}
  \Omega_{\mathrm{scattering}}^{\mathrm{pol}}\left(p',\eps',z\right)
   \wedge
  \Omega_{\mathrm{scattering}}^{\mathrm{pol}}\left(p'',\eps'',z\right),
\eq
which proves the claim.

\subsubsection{Examples}

Let us look at a few examples.
We consider again $\mathcal M_{0,n}^\pi$ with $\pi=(1,2,...,n)$.
We fix $z_1=0$, $z_{n-1}=1$ and $z_n=\infty$.
We look at the scattering form $\Omega_{\mathrm{scattering}}^{\mathrm{pol}}$ for external gauge bosons.
The simplest case is $n=3$.
We have the $0$-form
\bq
\lefteqn{
 \Omega_{\mathrm{scattering}}^{\mathrm{pol}}\left(p,\eps,z\right)
 = } & & \nonumber \\
 & &
   \left( \eps_1 \cdot \eps_2 \right) \left( \eps_3 \cdot \left( p_1 - p_2 \right) \right)
 + \left( \eps_2 \cdot \eps_3 \right) \left( \eps_1 \cdot \left( p_2 - p_3 \right) \right)
 + \left( \eps_3 \cdot \eps_1 \right) \left( \eps_2 \cdot \left( p_3 - p_1 \right) \right).
\eq
Up to a factor $i$, this is the three-point amplitude or equivalently the three-point vertex.

The next case is $n=4$. Here we have
\bq
 \Omega_{\mathrm{scattering}}^{\mathrm{pol}}\left(p,\eps,z\right)
 = 
 \left[
  \frac{N^{\mathrm{BCJ}}_{\mathrm{comb}}\left((1,2,3,4)\right)}{u_{2,4} \left(u_{2,4}-1\right)}
  -
  \frac{N^{\mathrm{BCJ}}_{\mathrm{comb}}\left((1,3,2,4)\right)}{\left(u_{2,4}-1\right)}
 \right] du_{2,4},
\eq
with 
\bq
\lefteqn{
 N^{\mathrm{BCJ}}_{\mathrm{comb}}\left((1,2,3,4)\right)
 =
} & & \nonumber \\
 & &
   4 \left( p_1 \cdot \eps_2 \right) \left( p_4 \cdot \eps_3 \right) \left( \eps_1 \cdot \eps_4 \right)
 - 4 \left( p_1 \cdot \eps_2 \right) \left( p_3 \cdot \eps_4 \right) \left( \eps_1 \cdot \eps_3 \right)
 - 4 \left( p_2 \cdot \eps_1 \right) \left( p_4 \cdot \eps_3 \right) \left( \eps_2 \cdot \eps_4 \right)
 \nonumber \\
 & &
 + 4 \left( p_2 \cdot \eps_1 \right) \left( p_3 \cdot \eps_4 \right) \left( \eps_2 \cdot \eps_3 \right)
 - 4 \left[ \left( p_1 \cdot \eps_3 \right) \left( p_2 \cdot \eps_4 \right) 
          - \left( p_1 \cdot \eps_4 \right) \left( p_2 \cdot \eps_3 \right) \right] \left( \eps_1 \cdot \eps_2 \right)
 \nonumber \\
 & &
 - 4 \left[ \left( p_3 \cdot \eps_1 \right) \left( p_4 \cdot \eps_2 \right) 
          - \left( p_3 \cdot \eps_2 \right) \left( p_4 \cdot \eps_1 \right) \right] \left( \eps_3 \cdot \eps_4 \right)
 - 2 \left( p_1 \cdot p_2 \right) \left( \eps_1 \cdot \eps_3 \right) \left( \eps_2 \cdot \eps_4 \right)
 \nonumber \\
 & &
 + 2 \left( p_1 \cdot p_2 \right) \left( \eps_1 \cdot \eps_4 \right) \left( \eps_2 \cdot \eps_3 \right)
 + 2 \left( p_2 \cdot p_3 - p_1 \cdot p_3 \right) \left( \eps_1 \cdot \eps_2 \right) \left( \eps_3 \cdot \eps_4 \right),
\eq
and $N^{\mathrm{BCJ}}_{\mathrm{comb}}((1,3,2,4))$ is obtained from $N^{\mathrm{BCJ}}_{\mathrm{comb}}((1,2,3,4))$
by $2 \leftrightarrow 3$. 

Finally, let us consider the case $n=5$.
Here we have
\bq
\lefteqn{
 \Omega_{\mathrm{scattering}}^{\mathrm{pol}}\left(p,\eps,z\right)
 = 
 \left[
 \frac{N^{\mathrm{BCJ}}_{\mathrm{comb}}\left((1,2,3,4,5)\right)}{u_{2,5} \left(u_{2,5}-1\right) u_{3,5} \left(u_{3,5}-1\right)}
 - \frac{N^{\mathrm{BCJ}}_{\mathrm{comb}}\left((1,3,4,2,5)\right)}{\left(u_{3,5}-1\right) \left(u_{2,5} u_{3,5}-1\right)}
 \right.
 } & & 
 \nonumber \\
 & &
 \left.
 - \frac{N^{\mathrm{BCJ}}_{\mathrm{comb}}\left((1,4,2,3,5)\right)}{\left(u_{2,5}-1\right) \left(u_{2,5} u_{3,5}-1\right)}
 + \frac{N^{\mathrm{BCJ}}_{\mathrm{comb}}\left((1,4,3,2,5)\right)}{\left(u_{2,5}-1\right) \left(u_{3,5}-1\right)}
 - \frac{N^{\mathrm{BCJ}}_{\mathrm{comb}}\left((1,2,4,3,5)\right)}{u_{2,5} \left(u_{3,5}-1\right) \left(u_{2,5} u_{3,5}-1\right)}
 \right.
 \nonumber \\
 & & \left.
 - \frac{N^{\mathrm{BCJ}}_{\mathrm{comb}}\left((1,3,2,4,5)\right)}{\left(u_{2,5}-1\right) u_{3,5} \left(u_{2,5} u_{3,5}-1\right)}
 \right] du_{2,5} \wedge du_{3,5}.
\eq
We may rewrite this expression with the help of the Leinartas decomposition \cite{Leinartas:1978,Raichev:2012,Meyer:2016slj} as
\bq
\lefteqn{
 \Omega_{\mathrm{scattering}}^{\mathrm{pol}}\left(p,\eps,z\right)
 = 
 \left[
 \frac{N^{\mathrm{BCJ}}_{\mathrm{comb}}\left((1,2,3,4,5)\right)}{u_{2,5} u_{3,5}}
 + \frac{N^{\mathrm{BCJ}}_{\mathrm{comb}}\left((1,2,4,3,5)\right)-N^{\mathrm{BCJ}}_{\mathrm{comb}}\left((1,2,3,4,5)\right)}{u_{2,5} \left(u_{3,5}-1\right)}
 \right. } & & \nonumber \\
 & &
 \left.
 + \frac{N^{\mathrm{BCJ}}_{\mathrm{comb}}\left((1,3,2,4,5)\right)-N^{\mathrm{BCJ}}_{\mathrm{comb}}\left((1,2,3,4,5)\right)}{\left(u_{2,5}-1\right) u_{3,5}}
 + \frac{N^{\mathrm{BCJ}}_{\mathrm{comb}}\left((1,4,3,2,5)\right)+N^{\mathrm{BCJ}}_{\mathrm{comb}}\left((1,2,3,4,5)\right)}{\left(u_{2,5}-1\right) \left(u_{3,5}-1\right)}
 \right. \nonumber \\
 & & \left.
 - \frac{N^{\mathrm{BCJ}}_{\mathrm{comb}}\left((1,4,2,3,5)\right)+N^{\mathrm{BCJ}}_{\mathrm{comb}}\left((1,3,2,4,5)\right)}{\left(u_{2,5}-1\right) \left(u_{2,5} u_{3,5}-1\right)}
 - \frac{N^{\mathrm{BCJ}}_{\mathrm{comb}}\left((1,3,4,2,5)\right)+N^{\mathrm{BCJ}}_{\mathrm{comb}}\left((1,2,4,3,5)\right)}{\left(u_{3,5}-1\right) \left(u_{2,5} u_{3,5}-1\right)}
 \right. \nonumber \\
 & & \left.
 - \frac{N^{\mathrm{BCJ}}_{\mathrm{comb}}\left((1,2,4,3,5)\right)+N^{\mathrm{BCJ}}_{\mathrm{comb}}\left((1,3,2,4,5)\right)}{\left(u_{2,5} u_{3,5}-1\right)}
 \right] du_{2,5} \wedge du_{3,5}.
\eq
All occuring BCJ-numerators may be obtained from the BCJ-numerator $N^{\mathrm{BCJ}}_{\mathrm{comb}}((1,2,3,4,5))$
by a suitable substitution of the indices.
The explicit expression for $N^{\mathrm{BCJ}}_{\mathrm{comb}}((1,2,3,4,5))$ is rather long and 
not reproduced here.
It is obtained in a straightforward way from the effective Lagrangian in eq.~(\ref{YM_vers2}).
Let us however point out that $N^{\mathrm{BCJ}}_{\mathrm{comb}}((1,2,3,4,5))$ is not unique.
This is related to the free parameter $a$ in eq.~(\ref{freedom_five}).

\subsection{Summary on the factorisation of the residues}

At the end of this section, let us summarise the factorisation properties of the residues of the scattering forms
$\Omega_{\mathrm{scattering}}^{\mathrm{cyclic}}(\sigma,z)$ and $\Omega_{\mathrm{scattering}}^{\mathrm{pol}}(p,\eps,z)$.
On $\mathcal M_{0,n}^\pi$ with $\pi=(1,2,...,n)$ we have in the limit $u_{i_0,n} \rightarrow 0$
\bq
 \mathrm{Res}_{Y} \Omega_{\mathrm{scattering}}^{\mathrm{cyclic}}\left(\sigma,z\right)
 & = &
 \left\{
 \begin{array}{ll}
  \left(-1\right)^{i_0-1}
  \Omega_{\mathrm{scattering}}^{\mathrm{cyclic}}\left(\sigma',z\right)
   \wedge
  \Omega_{\mathrm{scattering}}^{\mathrm{cyclic}}\left(\sigma'',z\right)
 & \mbox{if} \;\; \sigma \sim_{(i_0,n)} \pi, \\
 0 & \mbox{otherwise}. \\
 \end{array}
 \right. \nonumber \\
 \mathrm{Res}_{Y} \Omega_{\mathrm{scattering}}^{\mathrm{pol}}\left(p,\eps,z\right)
 & = &
  \left(-1\right)^{i_0-1}
  \sum\limits_{f, \; \lambda} 
  \Omega_{\mathrm{scattering}}^{\mathrm{pol}}\left(p',\eps',z\right)
   \wedge
  \Omega_{\mathrm{scattering}}^{\mathrm{pol}}\left(p'',\eps'',z\right),
\eq
where $Y$ denotes the hypersurface $u_{i_0,n}=0$.

\section{Conclusions}
\label{sect:conclusions}

In this paper we studied the properties of the scattering forms
$\Omega_{\mathrm{scattering}}^{\mathrm{cyclic}}$ and $\Omega_{\mathrm{scattering}}^{\mathrm{pol}}$.
These are two differential $(n-3)$-forms defined on the compactifiaction
$\overline{\mathcal M}_{0,n}$ of the moduli space of a Riemann surface of genus $0$
with $n$ marked points.
The scattering forms are cocycles.
The scattering equations define a one-form $\eta$ and Mizera \cite{Mizera:2017rqa}
has shown recently
that the scattering amplitudes are given as intersection numbers of the scattering forms
twisted by the one-form $\eta$.
With the two scattering forms 
$\Omega_{\mathrm{scattering}}^{\mathrm{cyclic}}$ and $\Omega_{\mathrm{scattering}}^{\mathrm{pol}}$
at hand we obtain, depending on the combination we take, the scattering amplitudes
within the bi-adjoint scalar theory, Yang-Mills theory and gravity.

In this paper we investigated the scattering forms in more detail.
We studied them on the complete $(n-3)$-dimensional space $\overline{\mathcal M}_{0,n}$,
not just on the zero-dimensional sub-variety defined by the scattering equations.

The scattering forms have some remarkable properties, 
given at the beginning of section~\ref{sect:scattering_forms}.
We have shown that the only singularities of the scattering forms 
are on the divisor $\overline{\mathcal M}_{0,n} \backslash {\mathcal M}_{0,n}$,
that all singularities are logarithmic
and that the residues at the singularities factorise into two scattering forms of lower points.
These properties provide a direct bridge from the scattering forms, obtained from the CHY representation,
to recent ideas involving associahedra and 
amplituhedra \cite{Arkani-Hamed:2017tmz,Arkani-Hamed:2013jha,Arkani-Hamed:2013kca,Arkani-Hamed:2014dca,Bai:2014cna,Franco:2014csa,Bern:2015ple}.
It is probably fair to say that we now have a clear geometric picture of tree-level amplitudes
within the bi-adjoint scalar theory, Yang-Mills theory and gravity for any number of
external particles $n$.

We expect these ideas to be fruitful for a wider set of theories \cite{Cachazo:2014xea,delaCruz:2015dpa,delaCruz:2015raa,delaCruz:2016wbr,Cachazo:2014nsa,Stieberger:2016lng,Nandan:2016pya,delaCruz:2016gnm,Fu:2017uzt,Teng:2017tbo,Chiodaroli:2017ngp,Johansson:2014zca,Johansson:2015oia}.
Another promising and interesting direction is to explore these ideas beyond tree-level amplitudes.

\subsection*{Acknowledgements}

A.K. is grateful for financial support from the research training group GRK 1581.

\begin{appendix}

\section{The reduced Pfaffian}
\label{sect:CHY_integrands}

\subsection{Definition}

In this appendix we collect the original definition of the polarisation factor $E(p,\eps,z)$
in terms of a reduced Pfaffian.
We start from a $(2n)\times(2n)$ anti-symmetric matrix $\Psi$ 
\bq
 \Psi 
 & = &
 \left( \begin{array}{cc}
 A & - C^T \\
 C & B \\
 \end{array} \right)
\eq
with
\bq
 A_{ab}
 = 
 \left\{ \begin{array}{cc}
 \frac{2 p_a \cdot p_b}{z_a-z_b} & a \neq b, \\
 0 & a = b, \\
 \end{array} \right.
 & &
 B_{ab}
 = 
 \left\{ \begin{array}{cc}
 \frac{2 \eps_a \cdot \eps_b}{z_a-z_b} & a \neq b, \\
 0 & a = b, \\
 \end{array} \right.
\eq
and
\bq
 C_{ab}
 & = &
 \left\{ \begin{array}{cc}
 \frac{2 \eps_a \cdot p_b}{z_a-z_b} & a \neq b, \\
 - \sum\limits_{j=1, j\neq a}^n \frac{2 \eps_a \cdot p_j}{z_a-z_j}  & a = b. \\
 \end{array} \right.
\eq
Let $1 \le i < j \le n$.
One denotes by $\Psi^{ij}_{ij}$ the $(2n-2)\times(2n-2)$-matrix 
where the rows and columns $i$ and $j$ of $\Psi$ have been deleted.
$\Psi^{ij}_{ij}$ has a non-vanishing Pfaffian.
The original definition of the polarisation factor reads \cite{Cachazo:2013gna,Cachazo:2013hca,Cachazo:2013iea}
\bq
 E\left(p,\eps,z\right)
 & = &
 \frac{\left(-1\right)^{i+j}}{2 \left(z_i-z_j\right)} \mathrm{Pf} \; \Psi^{ij}_{ij}.
\eq
This expression is independent of the choice of $i$ and $j$ 
if the momenta $p$ are on-shell,
the polarisation vectors $\eps$ are transverse and
the variables $z$ are solutions of the scattering equations.

\subsection{A counter-example}
\label{counter_example}

In this appendix we show that the reduced Pfaffian is not suited to define the polarisation scattering form.
Let us start from
\bq
 \frac{\left(-1\right)^{i+j} \mathrm{Pf} \; \Psi^{ij}_{ij}}{2 \left(z_i-z_j\right)} 
 \;
 \frac{d^nz}{d\omega}.
\eq
The first problem we face is that this expression is 
not independent of the deleted rows and columns $i$ and $j$ as soon as we go away 
from the solutions of the scattering equations.
We may overcome this problem by averaging over all possible choices of $i$ and $j$.
Thus we are tempted to consider
\bq
 \Omega^{\mathrm{try}}\left(p,\eps,z\right)
 & = &
 \frac{2}{n\left(n-1\right)}
 \sum\limits_{i<j}
 \frac{\left(-1\right)^{i+j} \mathrm{Pf} \; \Psi^{ij}_{ij}}{2 \left(z_i-z_j\right)} 
 \;
 \frac{d^nz}{d\omega}.
\eq
Let us specialise to the case $n=4$. One may show that $\Omega^{\mathrm{try}}$ is 
$\mathrm{PSL}(2,{\mathbb C})$-invariant.
We are considering $\mathcal M_{0,4}^\pi$ with $\pi=(1,2,3,4)$.
With the gauge choice $z_1=0$, $z_3=1$ and $z_4=\infty$ one finds
\bq
 \Omega^{\mathrm{try}}\left(p,\eps,z\right)
 & = &
 - \frac{2}{3} \left( \eps_1 \cdot \eps_2 \right) \left( \eps_3 \cdot \eps_4 \right) s_{12} \frac{du_{2,4}}{u_{2,4}^2}
 + ...,
\eq
where the dots denote terms less singular at $u_{2,4}=0$.
Thus $\Omega^{\mathrm{try}}$ has higher poles and does not satisfy the property of having only logarithmic singularities on the 
divisor $\overline{\mathcal M}_{0,n} \backslash {\mathcal M}_{0,n}$.

\section{Reduction of the numerators}
\label{sect:BCJ_numerator_reduction}

Let $\sigma=(1,2,...,n)$ be a cyclic order and $G \in {\mathcal T}_n(\sigma)$ a graph with this cyclic order.
In this appendix we show that
\bq
\label{graph_decomposition}
 N^{\mathrm{BCJ}}(G) & = & \hat{N}^{\mathrm{BCJ}}(G),
\eq
where $\hat{N}^{\mathrm{BCJ}}(G)$ is defined by
\bq
 \hat{N}^{\mathrm{BCJ}}(G)
 & = &
 \sum\limits_{\tilde{\sigma} \in \mathrm{CO}(G)}
 \sum\limits_{\kappa \in S_{n-2}^{(1,n)}}
 \delta_{\tilde{\sigma},\kappa}
 \left(-1\right)^{n_{\mathrm{flip}}(\sigma,\kappa)}
 N^{\mathrm{BCJ}}_{\mathrm{comb}}\left(\kappa\right).
\eq
Eq.~(\ref{graph_decomposition}) gives the reduction of an arbitrary BCJ-numerator into the basis
of multi-peripheral BCJ-numerators by repeated use of eq.~(\ref{STU_relation}).
Let us first look at an example.
\begin{figure}
\begin{center}
\includegraphics[scale=1.0]{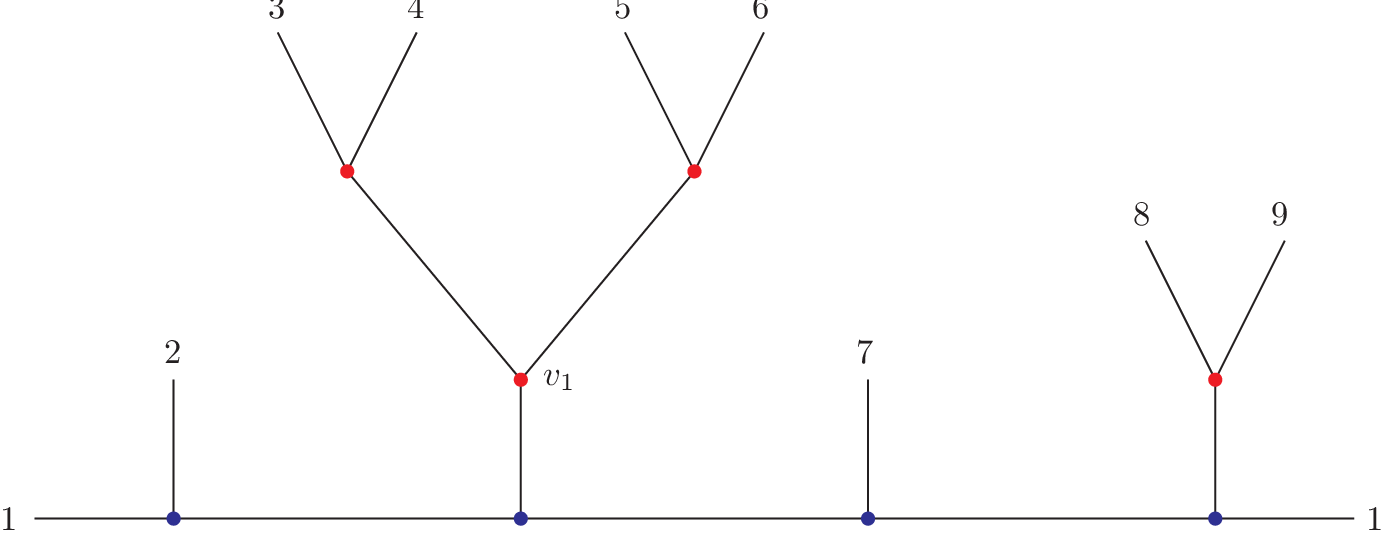}
\end{center}
\caption{
An example of a generic graph with the cyclic order $(1,2,...,10)$.
Vertices on the line connecting particle $1$ with $10$ are drawn in blue, all other vertices are
drawn in red.
}
\label{fig_bcj_reduction}
\end{figure}
Fig.~(\ref{fig_bcj_reduction}) shows an example of a generic graph with the cyclic order $(1,2,...,10)$.
\begin{figure}
\begin{center}
\includegraphics[scale=0.5]{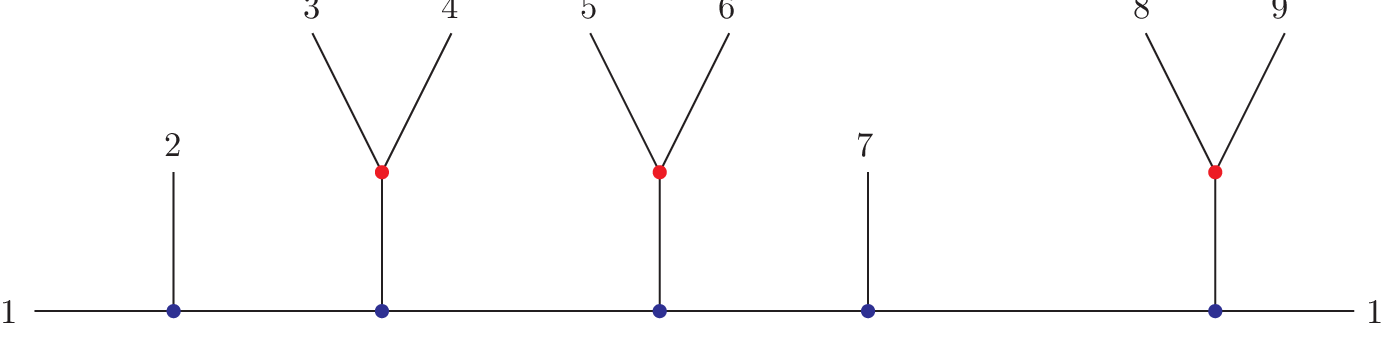}
\hspace*{10mm}
\includegraphics[scale=0.5]{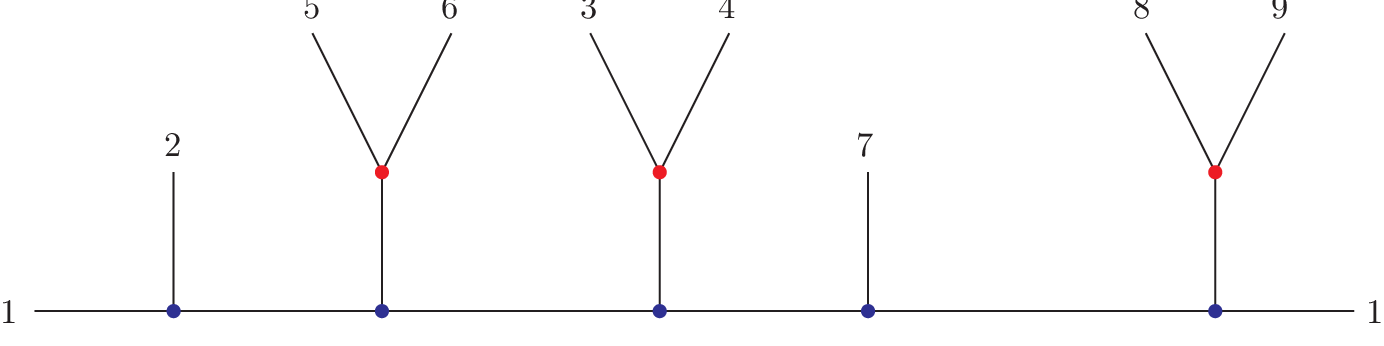}
\end{center}
\caption{
Reduction of the graph from fig.~(\ref{fig_bcj_reduction}): A single application
of the STU-relation at the vertex $v_1$ yields the two graphs shown in this figure.
The left graph occurs with a plus sign, the right graph with a minus sign.
}
\label{fig_bcj_reduction_2}
\end{figure}
If we apply the STU-relation at the vertex $v_1$, we obtain the two graphs shown in fig.~(\ref{fig_bcj_reduction_2}).
The left graph has the cyclic order $(1,2,3,4,5,6,7,8,9,10)$ and comes with a plus sign, 
the right graph has the cyclic order $(1,2,5,6,3,4,7,8,9,10)$ and comes with a minus sign.
These are exactly the cyclic orders (with the correct sign) we obtain from swapping the branches at
the vertex $v_1$.
We have a one-to-one correspondence between the terms occurring in the decomposition of
$N^{\mathrm{BCJ}}(G)$ into a multi-peripheral basis and the cyclic orders obtained by swapping in all
possible ways the vertices indicated by red (but not the ones shown in blue)
in fig.~(\ref{fig_bcj_reduction}).
Let us adopt the convention, that in any cyclic order leg $10$ occurs in the last place.
Then, swapping the branches at a red vertex will keep leg $1$ in the first place, whereas 
swapping the branches at a blue vertex will give a cyclic order where leg $1$ is not in the first place.
We may therefore sum over all cyclic orders compatible with $G$ and veto the ones which correspond
to a swap at a blue vertex.
The ones which correspond to swaps at red vertices only are the cyclic
orders $\kappa \in S_{n-2}^{(1,n)}$.
We thus arrive at
\bq
\label{proof_final_formula}
 N^{\mathrm{BCJ}}(G) 
 & = &
 \sum\limits_{\tilde{\sigma} \in \mathrm{CO}(G)}
 \sum\limits_{\kappa \in S_{n-2}^{(1,n)}}
 \delta_{\tilde{\sigma},\kappa}
 \left(-1\right)^{n_{\mathrm{flip}}(\sigma,\kappa)}
 N^{\mathrm{BCJ}}_{\mathrm{comb}}\left(\kappa\right),
\eq
where $\delta_{\tilde{\sigma},\kappa}$ selects exactly the cyclic orders, which are obtained 
from swapping branches at the red vertices only.

\section{Permutation invariance of the polarisation factor}
\label{sect:perm_invariance}

In this appendix we show the permutation invariance of the polarisation factor.
We set
\bq
 E_{1,n}
 & = &
 \sum\limits_{\kappa \in S_{n-2}^{(1,n)}}
 C\left(\kappa,z\right)
 \;
 N^{\mathrm{BCJ}}_{\mathrm{comb}}\left(\kappa\right),
 \nonumber \\
 E_{i,j}
 & = &
 \sum\limits_{\kappa \in S_{n-2}^{(i,j)}}
 C\left(\kappa,z\right)
 \;
 N^{\mathrm{BCJ}}_{\mathrm{comb}}\left(\kappa\right).
\eq
We would like to show that
\bq
 E_{i,j}
 & = &
 E_{1,n}.
\eq
We may split this into two steps, by first establishing
$E_{i,j}=E_{1,j}$ and then in a second step $E_{1,j}=E_{1,n}$.
The proof for the step, which exchanges $i \leftrightarrow 1$ is similar to the proof for the step,
which exchanges $j \leftrightarrow n$, therefore it suffices to discuss one case.
We show
\bq
 E_{1,j}
 & = &
 E_{1,n}.
\eq
$E_{1,j}$ is given by
\bq
 E_{1,j}
 & = &
 \sum\limits_{\kappa' \in S_{n-2}^{(1,j)}}
 C\left(\kappa',z\right)
 \;
 N^{\mathrm{BCJ}}_{\mathrm{comb}}\left(\kappa'\right),
\eq
where $\kappa'$ is of the form $(1,\alpha',n,\beta',j)$ with $\alpha' \cup \beta' = \{2,...,n-1\} \backslash \{j\}$
and $\alpha' \cap \beta' = \emptyset$.
We may replace the sum over $\kappa'$ by a sum over all possible choices for $\alpha'$ and $\beta'$.
Thus we write
\bq
 E_{1,j}
 & = &
 \sum\limits_{\alpha',\beta'}
 C\left((1,\alpha',n,\beta',j),z\right)
 \;
 N^{\mathrm{BCJ}}_{\mathrm{comb}}\left((1,\alpha',n,\beta',j)\right).
\eq
Let $G'$ be the multi-peripheral graph corresponding to the BCJ-numerator $N^{\mathrm{BCJ}}_{\mathrm{comb}}((1,\alpha',n,\beta',j))$.
We denote by $F'$ the graph obtained from the graph $G'$ by swapping the branches at the vertex where leg $n$ is attached.
\begin{figure}
\begin{center}
\includegraphics[scale=0.6]{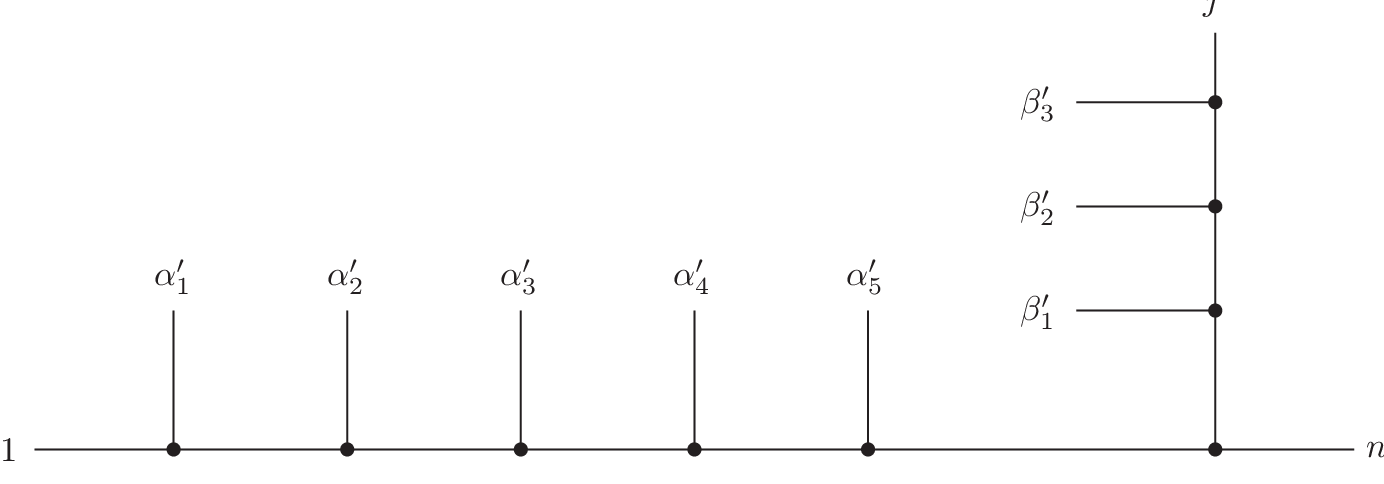}
\end{center}
\caption{
The graph $F'$ obtained from the graph underlying the numerator $N^{\mathrm{BCJ}}_{\mathrm{comb}}((1,\alpha',n,\beta',j))$ by swapping the branches at the vertex where leg $n$ is attached.
}
\label{fig_perm_graph}
\end{figure}
The graph $F'$ has the cyclic order $\kappa_F'=(1,\alpha',\beta',j,n)$.
An example is shown in fig.~(\ref{fig_perm_graph}).
We have
\bq
 N^{\mathrm{BCJ}}\left(G'\right) 
 & = &
 - N^{\mathrm{BCJ}}\left(F'\right).
\eq
Clearly
\bq
 \mathrm{CO}\left(G'\right) & = & \mathrm{CO}\left(F'\right)
\eq
and for any cyclic order $\sigma$
\bq
 \left(-1\right)^{n_{\mathrm{flip}}(\kappa',\sigma)}
 & = &
 -
 \left(-1\right)^{n_{\mathrm{flip}}(\kappa_F',\sigma)},
\eq
since $F'$ differs from $G'$ by exactly one swap.
From eq.~(\ref{proof_final_formula}) it follows that
\bq
 N^{\mathrm{BCJ}}\left(G'\right) 
 & = &
 \sum\limits_{\tilde{\sigma} \in \mathrm{CO}(G')}
 \sum\limits_{\kappa \in S_{n-2}^{(1,n)}}
 \delta_{\tilde{\sigma},\kappa}
 \left(-1\right)^{n_{\mathrm{flip}}(\kappa',\kappa)}
 N^{\mathrm{BCJ}}_{\mathrm{comb}}\left(\kappa\right).
\eq
Thus
\bq
 E_{1,j}
 & = &
 \sum\limits_{\kappa \in S_{n-2}^{(1,n)}}
 N^{\mathrm{BCJ}}_{\mathrm{comb}}\left(\kappa\right)
 \sum\limits_{\kappa' \in S_{n-2}^{(1,j)}}
 C\left(\kappa',z\right)
 \sum\limits_{\tilde{\sigma} \in \mathrm{CO}(G')}
 \delta_{\tilde{\sigma},\kappa}
 \left(-1\right)^{n_{\mathrm{flip}}(\kappa',\kappa)}.
\eq
This equals $E_{1,n}$ if
\bq
\label{relation_cyclic_factors}
 C\left(\kappa,z\right)
 & = &
 \sum\limits_{\alpha',\beta'}
 C\left(\kappa',z\right)
 \sum\limits_{\tilde{\sigma} \in \mathrm{CO}(G')}
 \delta_{\tilde{\sigma},\kappa}
 \left(-1\right)^{n_{\mathrm{flip}}(\kappa',\kappa)},
\eq
with $\kappa'=(1,\alpha',n,\beta',j)$.
With the help of the Kleiss-Kuijf relations we may express $C(\kappa',z)$
as
\bq
 C\left(\kappa',z\right)
 & = &
 \left(-1\right)^{1+|\beta'|}
 \sum\limits_{\sigma \in (1, \alpha' \; \shuffle \; (j,\beta'^T),n)}
 C\left(\sigma,z\right),
\eq
where $|\beta'|$ denotes the number of elements of $\beta'$.
Thus the right-hand side of eq.~(\ref{relation_cyclic_factors}) equals
\bq
 \mbox{r.h.s.}
 & = &
 \sum\limits_{\alpha',\beta'}
 \left(-1\right)^{1+|\beta'|}
 \sum\limits_{\sigma \in (1, \alpha' \; \shuffle \; (j,\beta'^T),n)}
 C\left(\sigma,z\right)
 \sum\limits_{\tilde{\sigma} \in \mathrm{CO}(G')}
 \delta_{\tilde{\sigma},\kappa}
 \left(-1\right)^{n_{\mathrm{flip}}(\kappa',\kappa)}
 \\
 & = & 
 \sum\limits_{\kappa'' \in S_{n-2}^{(1,n)}}
 C\left(\kappa'',z\right)
 \sum\limits_{\alpha',\beta'}
 \left(-1\right)^{1+|\beta'|}
 \sum\limits_{\sigma \in (1, \alpha' \; \shuffle \; (j,\beta'^T),n)}
 \delta_{\kappa'',\sigma}
 \sum\limits_{\tilde{\sigma} \in \mathrm{CO}(G')}
 \delta_{\tilde{\sigma},\kappa}
 \left(-1\right)^{n_{\mathrm{flip}}(\kappa',\kappa)}.
 \nonumber
\eq
Therefore it remains to show
\bq
\label{relation_combinatorics}
 \delta_{\kappa,\kappa''}
 & = &
 \sum\limits_{\kappa' = (1,\alpha',n,\beta',j) \in S_{n-2}^{(1,j)}}
 \;\;\;
 \sum\limits_{\sigma \in (1, \alpha' \; \shuffle \; (j,\beta'^T),n)}
 \;\;\;
 \sum\limits_{\tilde{\sigma} \in \mathrm{CO}(G')}
 \delta_{\kappa'',\sigma}
 \delta_{\tilde{\sigma},\kappa}
 \left(-1\right)^{1+|\beta'|+n_{\mathrm{flip}}(\kappa',\kappa)}.
 \;\;\;\;\;\;\;\;\;
\eq
Let us first discuss the case $\kappa=\kappa''$.
Without loss of generality we may assume $\kappa=\kappa''=(1,2,...,n)$.
The right-hand side of eq.~(\ref{relation_combinatorics}) simplifies to
\bq
 \mbox{r.h.s.}
 & = &
 \sum\limits_{\kappa' = (1,\alpha',n,\beta',j) \in S_{n-2}^{(1,j)}}
 \;\;\;
 \sum\limits_{\sigma \in (1, \alpha' \; \shuffle \; (j,\beta'^T),n)}
 \;\;\;
 \sum\limits_{\tilde{\sigma} \in \mathrm{CO}(G')}
 \delta_{\kappa,\sigma}
 \delta_{\sigma,\tilde{\sigma}}
 \left(-1\right)^{1+|\beta'|+n_{\mathrm{flip}}(\kappa',\kappa)}.
 \;\;\;\;\;\;
\eq
The Kronecker delta $\delta_{\sigma,\tilde{\sigma}}$ selects from the sums over $\sigma$ and $\tilde{\sigma}$
the term $\sigma=\tilde{\sigma}=(1,\alpha',j,\beta'^T,n)$.
In addition we must have
$\kappa=\sigma$, i.e. $\alpha'=(2,...,j-1)$ and $\beta'^T=(j+1,...,n-1)$.
Therefore also the sum over $\kappa'$ reduces to one term.
For $\kappa'=(1,2,...,j-1,n,n-1,...,j+1,j)$ we have
\bq
 n_{\mathrm{flip}}(\kappa',\kappa)
 & = & n-j.
\eq
We obtain
for the right-hand side of eq.~(\ref{relation_combinatorics})
\bq
 \mbox{r.h.s.}
 & = &
 \left(-1\right)^{1+|\beta'|+n_{\mathrm{flip}}(\kappa',\kappa)}
 \;\; = \;\;
 \left(-1\right)^{1+\left(n-1-j\right)+\left(n-j\right)}
 \;\; = \;\;
 1.
\eq
Let us now discuss for $\kappa,\kappa'' \in S_{n-2}^{(1,n)}$
the case $\kappa \neq \kappa''$.
This requires $n \ge 4$.
Without loss of generality we may take $\kappa''=(1,2,...,n)$ and $j=n-1$.
The Kronecker delta $\delta_{\kappa'',\sigma}$ enforces then $\beta'=\emptyset$ and $\alpha'=(2,...,n-2)$.
We thus have $\kappa'=(1,2,...,n-2,n,n-1)$.
The graph $G'$ is then the multi-peripheral graph
\bq
\begin{picture}(210,60)(0,0)
\Line(10,10)(200,10)
\Vertex(35,10){2}
\Vertex(65,10){2}
\Vertex(95,10){2}
\Vertex(145,10){2}
\Vertex(175,10){2}
\Line(35,10)(35,40)
\Line(65,10)(65,40)
\Line(95,10)(95,40)
\Line(145,10)(145,40)
\Line(175,10)(175,40)
\Text(5,10)[r]{$1$}
\Text(35,45)[b]{$2$}
\Text(65,45)[b]{$3$}
\Text(120,25)[b]{$...$}
\Text(145,45)[b]{$n-2$}
\Text(175,45)[b]{$n$}
\Text(205,10)[l]{$n-1$}
\end{picture}
\eq
and the graph $F'$ is the multi-peripheral graph
\bq
\begin{picture}(210,60)(0,0)
\Line(10,10)(200,10)
\Vertex(35,10){2}
\Vertex(65,10){2}
\Vertex(95,10){2}
\Vertex(145,10){2}
\Vertex(175,10){2}
\Line(35,10)(35,40)
\Line(65,10)(65,40)
\Line(95,10)(95,40)
\Line(145,10)(145,40)
\Line(175,10)(175,40)
\Text(5,10)[r]{$1$}
\Text(35,45)[b]{$2$}
\Text(65,45)[b]{$3$}
\Text(120,25)[b]{$...$}
\Text(145,45)[b]{$n-2$}
\Text(175,45)[b]{$n-1$}
\Text(205,10)[l]{$n$.}
\end{picture}
\eq
The only cyclic order of $\mathrm{CO}(G')$, which is also an element of $S_{n-2}^{(1,n)}$ is
\bq
 \left(1,2,...,n\right).
\eq
Thus, the Kronecker delta $\delta_{\tilde{\sigma},\kappa}$ enforces
$\kappa=\tilde{\sigma}=(1,2,...,n)$. Thus $\kappa=\kappa''$. But this is a contradiction
to the assumption $\kappa \neq \kappa''$.
Therefore all terms on the right-hand side of eq.~(\ref{relation_combinatorics}) 
vanish due to the Kronecker deltas
and the right-hand side of eq.~(\ref{relation_combinatorics}) yields zero
for $\kappa \neq \kappa''$.
This completes the proof of the permutation invariance of 
the polarisation factor.

\end{appendix}

\bibliography{/home/stefanw/notes/biblio}

\providecommand{\href}[2]{#2}\begingroup\raggedright\begin{thebibliography}{10}

\bibitem{Cachazo:2013gna}
F.~Cachazo, S.~He and E.~Y. Yuan, \emph{{Scattering equations and
  Kawai-Lewellen-Tye orthogonality}},
  \href{https://doi.org/10.1103/PhysRevD.90.065001}{\emph{Phys.Rev.} {\bfseries
  D90} (2014) 065001}, [\href{https://arxiv.org/abs/1306.6575}{{\ttfamily
  1306.6575}}].

\bibitem{Cachazo:2013hca}
F.~Cachazo, S.~He and E.~Y. Yuan, \emph{{Scattering of Massless Particles in
  Arbitrary Dimensions}},
  \href{https://doi.org/10.1103/PhysRevLett.113.171601}{\emph{Phys.Rev.Lett.}
  {\bfseries 113} (2014) 171601},
  [\href{https://arxiv.org/abs/1307.2199}{{\ttfamily 1307.2199}}].

\bibitem{Cachazo:2013iea}
F.~Cachazo, S.~He and E.~Y. Yuan, \emph{{Scattering of Massless Particles:
  Scalars, Gluons and Gravitons}},
  \href{https://doi.org/10.1007/JHEP07(2014)033}{\emph{JHEP} {\bfseries 1407}
  (2014) 033}, [\href{https://arxiv.org/abs/1309.0885}{{\ttfamily 1309.0885}}].

\bibitem{Arkani-Hamed:2017tmz}
N.~Arkani-Hamed, Y.~Bai and T.~Lam, \emph{{Positive Geometries and Canonical
  Forms}},  \href{https://arxiv.org/abs/1703.04541}{{\ttfamily 1703.04541}}.

\bibitem{Mizera:2017rqa}
S.~Mizera, \emph{{Scattering Amplitudes from Intersection Theory}},
  \href{https://arxiv.org/abs/1711.00469}{{\ttfamily 1711.00469}}.

\bibitem{Mizera:2017cqs}
S.~Mizera, \emph{{Combinatorics and Topology of Kawai-Lewellen-Tye Relations}},
  \href{https://doi.org/10.1007/JHEP08(2017)097}{\emph{JHEP} {\bfseries 08}
  (2017) 097}, [\href{https://arxiv.org/abs/1706.08527}{{\ttfamily
  1706.08527}}].

\bibitem{Cachazo:2016ror}
F.~Cachazo, S.~Mizera and G.~Zhang, \emph{{Scattering Equations: Real Solutions
  and Particles on a Line}},
  \href{https://doi.org/10.1007/JHEP03(2017)151}{\emph{JHEP} {\bfseries 03}
  (2017) 151}, [\href{https://arxiv.org/abs/1609.00008}{{\ttfamily
  1609.00008}}].

\bibitem{Weinzierl:2014vwa}
S.~Weinzierl, \emph{{On the solutions of the scattering equations}},
  \href{https://doi.org/10.1007/JHEP04(2014)092}{\emph{JHEP} {\bfseries 1404}
  (2014) 092}, [\href{https://arxiv.org/abs/1402.2516}{{\ttfamily 1402.2516}}].

\bibitem{Kalousios:2013eca}
C.~Kalousios, \emph{{Massless scattering at special kinematics as Jacobi
  polynomials}},
  \href{https://doi.org/10.1088/1751-8113/47/21/215402}{\emph{J.Phys.}
  {\bfseries A47} (2014) 215402},
  [\href{https://arxiv.org/abs/1312.7743}{{\ttfamily 1312.7743}}].

\bibitem{Arkani-Hamed:2017mur}
N.~Arkani-Hamed, Y.~Bai, S.~He and G.~Yan, \emph{{Scattering Forms and the
  Positive Geometry of Kinematics, Color and the Worldsheet}},
  \href{https://arxiv.org/abs/1711.09102}{{\ttfamily 1711.09102}}.

\bibitem{Tolotti:2013caa}
M.~Tolotti and S.~Weinzierl, \emph{{Construction of an effective Yang-Mills
  Lagrangian with manifest BCJ duality}},
  \href{https://doi.org/10.1007/JHEP07(2013)111}{\emph{JHEP} {\bfseries 07}
  (2013) 111}, [\href{https://arxiv.org/abs/1306.2975}{{\ttfamily 1306.2975}}].

\bibitem{Draggiotis:1998gr}
P.~Draggiotis, R.~H.~P. Kleiss and C.~G. Papadopoulos, \emph{On the computation
  of multigluon amplitudes}, {\emph{Phys. Lett.} {\bfseries B439} (1998)
  157--164}, [\href{https://arxiv.org/abs/hep-ph/9807207}{{\ttfamily
  hep-ph/9807207}}].

\bibitem{Duhr:2006iq}
C.~Duhr, S.~Hoche and F.~Maltoni, \emph{Color-dressed recursive relations for
  multi-parton amplitudes}, {\emph{JHEP} {\bfseries 08} (2006) 062},
  [\href{https://arxiv.org/abs/hep-ph/0607057}{{\ttfamily hep-ph/0607057}}].

\bibitem{Weinzierl:2016bus}
S.~Weinzierl, \emph{{Tales of 1001 Gluons}},
  \href{https://doi.org/10.1016/j.physrep.2017.01.004}{\emph{Phys. Rept.}
  {\bfseries 676} (2017) 1--101},
  [\href{https://arxiv.org/abs/1610.05318}{{\ttfamily 1610.05318}}].

\bibitem{Bern:2008qj}
Z.~Bern, J.~J.~M. Carrasco and H.~Johansson, \emph{{New Relations for
  Gauge-Theory Amplitudes}},
  \href{https://doi.org/10.1103/PhysRevD.78.085011}{\emph{Phys. Rev.}
  {\bfseries D78} (2008) 085011},
  [\href{https://arxiv.org/abs/0805.3993}{{\ttfamily 0805.3993}}].

\bibitem{Bern:2010ue}
Z.~Bern, J.~J.~M. Carrasco and H.~Johansson, \emph{{Perturbative Quantum
  Gravity as a Double Copy of Gauge Theory}},
  \href{https://doi.org/10.1103/PhysRevLett.105.061602}{\emph{Phys.Rev.Lett.}
  {\bfseries 105} (2010) 061602},
  [\href{https://arxiv.org/abs/1004.0476}{{\ttfamily 1004.0476}}].

\bibitem{Bern:2010yg}
Z.~Bern, T.~Dennen, Y.-t. Huang and M.~Kiermaier, \emph{{Gravity as the Square
  of Gauge Theory}},
  \href{https://doi.org/10.1103/PhysRevD.82.065003}{\emph{Phys.Rev.} {\bfseries
  D82} (2010) 065003}, [\href{https://arxiv.org/abs/1004.0693}{{\ttfamily
  1004.0693}}].

\bibitem{Griffiths:book}
P.~Griffiths and J.~Harris, \emph{Principles of Algebraic Geometry}.
\newblock John Wiley \& Sons, New York, 1994.

\bibitem{Abreu:2017ptx}
S.~Abreu, R.~Britto, C.~Duhr and E.~Gardi, \emph{{Cuts from residues: the
  one-loop case}}, \href{https://doi.org/10.1007/JHEP06(2017)114}{\emph{JHEP}
  {\bfseries 06} (2017) 114},
  [\href{https://arxiv.org/abs/1702.03163}{{\ttfamily 1702.03163}}].

\bibitem{Deligne:1969}
P.~Deligne and D.~Mumford, \emph{The irreducibility of the space of curves of
  given genus}, {\emph{Publ. Math. Inst. Hautes Études Sci.} {\bfseries 36}
  (1969) 75}.

\bibitem{Knudsen:1976}
F.~Knudsen and D.~Mumford, \emph{{The projectivity of the moduli space of
  stable curves I: Preliminaries on "det" and "Div"}}, {\emph{Math. Scand.}
  {\bfseries 39} (1976) 19}.

\bibitem{Knudsen:1983}
F.~Knudsen, \emph{{The projectivity of the moduli space of stable curves II:
  The stacks $M_{g,n}$}}, {\emph{Math. Scand.} {\bfseries 52} (1983) 161}.

\bibitem{Knudsen:1983a}
F.~Knudsen, \emph{{The projectivity of the moduli space of stable curves III:
  The line bundles on $M_{g,n}$, and a proof of the projectivity of
  $\overline{M}_{g,n}$ in characteristic $0$}}, {\emph{Math. Scand.} {\bfseries
  52} (1983) 200}.

\bibitem{Brown:2006}
F.~Brown, \emph{{Multiple zeta values and periods of moduli spaces
  $\overline{\mathcal M}_{0,n}$}}, {\emph{C. R. Acad. Sci. Paris} {\bfseries
  342} (2006) 949}.

\bibitem{Stasheff:1963a}
J.~D. Stasheff, \emph{{Homotopy associativity of $H$-spaces. I}}, {\emph{Trans.
  Amer. Math. Soc.} {\bfseries 108} (1963) 275}.

\bibitem{Stasheff:1963b}
J.~D. Stasheff, \emph{{Homotopy associativity of $H$-spaces. II}},
  {\emph{Trans. Amer. Math. Soc.} {\bfseries 108} (1963) 293}.

\bibitem{Devadoss:1998}
S.~Devadoss, \emph{{Tessellations of Moduli Spaces and the Mosaic Operad}},
  \href{https://arxiv.org/abs/math/9807010}{{\ttfamily math/9807010}}.

\bibitem{Devadoss:2004}
S.~Devadoss, \emph{{Combinatorial equivalence of real moduli spaces}},
  {\emph{Notices of the AMS} {\bfseries 51} (2004) 620},
  [\href{https://arxiv.org/abs/math-ph/0405011}{{\ttfamily math-ph/0405011}}].

\bibitem{Kleiss:1988ne}
R.~Kleiss and H.~Kuijf, \emph{{Multi - Gluon Cross-sections and Five Jet
  Production at Hadron Colliders}},
  \href{https://doi.org/10.1016/0550-3213(89)90574-9}{\emph{Nucl. Phys.}
  {\bfseries B312} (1989) 616--644}.

\bibitem{Arkani-Hamed:amplitudes2017}
N.~Arkani-Hamed, \emph{{Spacetime, QM and Positive Geometry}}, {\emph{presented
  at Amplitudes 2017} (2017) }.

\bibitem{Bai:amplitudes2017}
Y.~Bai, \emph{{Positive Geometries and Canonical Forms}}, {\emph{talk presented
  at Amplitudes 2017} (2017) }.

\bibitem{Du:2013sha}
Y.-J. Du, B.~Feng and C.-H. Fu, \emph{{The Construction of Dual-trace Factor in
  Yang-Mills Theory}},
  \href{https://doi.org/10.1007/JHEP07(2013)057}{\emph{JHEP} {\bfseries 07}
  (2013) 057}, [\href{https://arxiv.org/abs/1304.2978}{{\ttfamily 1304.2978}}].

\bibitem{Litsey:2013jfa}
S.~Litsey and J.~Stankowicz, \emph{{Kinematic numerators and a double-copy
  formula for $N$=4 super-Yang-Mills residues}},
  \href{https://doi.org/10.1103/PhysRevD.90.025013}{\emph{Phys. Rev.}
  {\bfseries D90} (2014) 025013},
  [\href{https://arxiv.org/abs/1309.7681}{{\ttfamily 1309.7681}}].

\bibitem{Lam:2016tlk}
C.~S. Lam and Y.-P. Yao, \emph{{Evaluation of the Cachazo-He-Yuan gauge
  amplitude}}, \href{https://doi.org/10.1103/PhysRevD.93.105008}{\emph{Phys.
  Rev.} {\bfseries D93} (2016) 105008},
  [\href{https://arxiv.org/abs/1602.06419}{{\ttfamily 1602.06419}}].

\bibitem{Bjerrum-Bohr:2016axv}
N.~E.~J. Bjerrum-Bohr, J.~L. Bourjaily, P.~H. Damgaard and B.~Feng,
  \emph{{Manifesting Color-Kinematics Duality in the Scattering Equation
  Formalism}}, \href{https://doi.org/10.1007/JHEP09(2016)094}{\emph{JHEP}
  {\bfseries 09} (2016) 094},
  [\href{https://arxiv.org/abs/1608.00006}{{\ttfamily 1608.00006}}].

\bibitem{Huang:2017ydz}
R.~Huang, Y.-J. Du and B.~Feng, \emph{{Understanding the Cancelation of Double
  Poles in the Pfaffian of CHY-formulism}},
  \href{https://doi.org/10.1007/JHEP06(2017)133}{\emph{JHEP} {\bfseries 06}
  (2017) 133}, [\href{https://arxiv.org/abs/1702.05840}{{\ttfamily
  1702.05840}}].

\bibitem{Du:2017kpo}
Y.-J. Du and F.~Teng, \emph{{BCJ numerators from reduced Pfaffian}},
  \href{https://doi.org/10.1007/JHEP04(2017)033}{\emph{JHEP} {\bfseries 04}
  (2017) 033}, [\href{https://arxiv.org/abs/1703.05717}{{\ttfamily
  1703.05717}}].

\bibitem{Gao:2017dek}
X.~Gao, S.~He and Y.~Zhang, \emph{{Labelled tree graphs, Feynman diagrams and
  disk integrals}},  \href{https://arxiv.org/abs/1708.08701}{{\ttfamily
  1708.08701}}.

\bibitem{Chen:2016fgi}
T.~Wang, G.~Chen, Y.-K.~E. Cheung and F.~Xu, \emph{{A differential operator for
  integrating one-loop scattering equations}},
  \href{https://doi.org/10.1007/JHEP01(2017)028}{\emph{JHEP} {\bfseries 01}
  (2017) 028}, [\href{https://arxiv.org/abs/1609.07621}{{\ttfamily
  1609.07621}}].

\bibitem{Chen:2017edo}
T.~Wang, G.~Chen, Y.-K.~E. Cheung and F.~Xu, \emph{{A Combinatoric Shortcut to
  Evaluate CHY-forms}},
  \href{https://doi.org/10.1007/JHEP06(2017)015}{\emph{JHEP} {\bfseries 06}
  (2017) 015}, [\href{https://arxiv.org/abs/1701.06488}{{\ttfamily
  1701.06488}}].

\bibitem{Chen:2017bug}
G.~Chen and T.~Wang, \emph{{BCJ Numerators from Differential Operator of
  Multidimensional Residue}},
  \href{https://arxiv.org/abs/1709.08503}{{\ttfamily 1709.08503}}.

\bibitem{Leinartas:1978}
E.~K. Leinartas, \emph{{Factorization of rational functions of several
  variables into partial fractions}}, {\emph{Izv. Vyssh. Uchebn. Zaved. Mat.}
  {\bfseries 22} (1978) 47}.

\bibitem{Raichev:2012}
A.~{Raichev}, \emph{{Leinartas's partial fraction decomposition}},
  \href{https://arxiv.org/abs/1206.4740}{{\ttfamily 1206.4740}}.

\bibitem{Meyer:2016slj}
C.~Meyer, \emph{{Transforming differential equations of multi-loop Feynman
  integrals into canonical form}},
  \href{https://doi.org/10.1007/JHEP04(2017)006}{\emph{JHEP} {\bfseries 04}
  (2017) 006}, [\href{https://arxiv.org/abs/1611.01087}{{\ttfamily
  1611.01087}}].

\bibitem{Arkani-Hamed:2013jha}
N.~Arkani-Hamed and J.~Trnka, \emph{{The Amplituhedron}},
  \href{https://doi.org/10.1007/JHEP10(2014)030}{\emph{JHEP} {\bfseries 10}
  (2014) 030}, [\href{https://arxiv.org/abs/1312.2007}{{\ttfamily 1312.2007}}].

\bibitem{Arkani-Hamed:2013kca}
N.~Arkani-Hamed and J.~Trnka, \emph{{Into the Amplituhedron}},
  \href{https://doi.org/10.1007/JHEP12(2014)182}{\emph{JHEP} {\bfseries 12}
  (2014) 182}, [\href{https://arxiv.org/abs/1312.7878}{{\ttfamily 1312.7878}}].

\bibitem{Arkani-Hamed:2014dca}
N.~Arkani-Hamed, A.~Hodges and J.~Trnka, \emph{{Positive Amplitudes In The
  Amplituhedron}}, \href{https://doi.org/10.1007/JHEP08(2015)030}{\emph{JHEP}
  {\bfseries 08} (2015) 030},
  [\href{https://arxiv.org/abs/1412.8478}{{\ttfamily 1412.8478}}].

\bibitem{Bai:2014cna}
Y.~Bai and S.~He, \emph{{The Amplituhedron from Momentum Twistor Diagrams}},
  \href{https://doi.org/10.1007/JHEP02(2015)065}{\emph{JHEP} {\bfseries 02}
  (2015) 065}, [\href{https://arxiv.org/abs/1408.2459}{{\ttfamily 1408.2459}}].

\bibitem{Franco:2014csa}
S.~Franco, D.~Galloni, A.~Mariotti and J.~Trnka, \emph{{Anatomy of the
  Amplituhedron}}, \href{https://doi.org/10.1007/JHEP03(2015)128}{\emph{JHEP}
  {\bfseries 03} (2015) 128},
  [\href{https://arxiv.org/abs/1408.3410}{{\ttfamily 1408.3410}}].

\bibitem{Bern:2015ple}
Z.~Bern, E.~Herrmann, S.~Litsey, J.~Stankowicz and J.~Trnka, \emph{{Evidence
  for a Nonplanar Amplituhedron}},
  \href{https://doi.org/10.1007/JHEP06(2016)098}{\emph{JHEP} {\bfseries 06}
  (2016) 098}, [\href{https://arxiv.org/abs/1512.08591}{{\ttfamily
  1512.08591}}].

\bibitem{Cachazo:2014xea}
F.~Cachazo, S.~He and E.~Y. Yuan, \emph{{Scattering Equations and Matrices:
  From Einstein To Yang-Mills, DBI and NLSM}},
  \href{https://doi.org/10.1007/JHEP07(2015)149}{\emph{JHEP} {\bfseries 07}
  (2015) 149}, [\href{https://arxiv.org/abs/1412.3479}{{\ttfamily 1412.3479}}].

\bibitem{delaCruz:2015dpa}
L.~de~la Cruz, A.~Kniss and S.~Weinzierl, \emph{{Proof of the fundamental BCJ
  relations for QCD amplitudes}},
  \href{https://doi.org/10.1007/JHEP09(2015)197}{\emph{JHEP} {\bfseries 09}
  (2015) 197}, [\href{https://arxiv.org/abs/1508.01432}{{\ttfamily
  1508.01432}}].

\bibitem{delaCruz:2015raa}
L.~de~la Cruz, A.~Kniss and S.~Weinzierl, \emph{{The CHY representation of
  tree-level primitive QCD amplitudes}}, {\emph{JHEP} {\bfseries 11} (2015)
  217}, [\href{https://arxiv.org/abs/1508.06557}{{\ttfamily 1508.06557}}].

\bibitem{delaCruz:2016wbr}
L.~de~la Cruz, A.~Kniss and S.~Weinzierl, \emph{{Double Copies of Fermions as
  Matter that Interacts Only Gravitationally}},
  \href{https://doi.org/10.1103/PhysRevLett.116.201601}{\emph{Phys. Rev. Lett.}
  {\bfseries 116} (2016) 201601},
  [\href{https://arxiv.org/abs/1601.04523}{{\ttfamily 1601.04523}}].

\bibitem{Cachazo:2014nsa}
F.~Cachazo, S.~He and E.~Y. Yuan, \emph{{Einstein-Yang-Mills Scattering
  Amplitudes From Scattering Equations}},
  \href{https://doi.org/10.1007/JHEP01(2015)121}{\emph{JHEP} {\bfseries 01}
  (2015) 121}, [\href{https://arxiv.org/abs/1409.8256}{{\ttfamily 1409.8256}}].

\bibitem{Stieberger:2016lng}
S.~Stieberger and T.~R. Taylor, \emph{{New relations for
  Einstein–Yang–Mills amplitudes}},
  \href{https://doi.org/10.1016/j.nuclphysb.2016.09.014}{\emph{Nucl. Phys.}
  {\bfseries B913} (2016) 151--162},
  [\href{https://arxiv.org/abs/1606.09616}{{\ttfamily 1606.09616}}].

\bibitem{Nandan:2016pya}
D.~Nandan, J.~Plefka, O.~Schlotterer and C.~Wen, \emph{{Einstein-Yang-Mills
  from pure Yang-Mills amplitudes}},
  \href{https://doi.org/10.1007/JHEP10(2016)070}{\emph{JHEP} {\bfseries 10}
  (2016) 070}, [\href{https://arxiv.org/abs/1607.05701}{{\ttfamily
  1607.05701}}].

\bibitem{delaCruz:2016gnm}
L.~de~la Cruz, A.~Kniss and S.~Weinzierl, \emph{{Relations for
  Einstein–Yang–Mills amplitudes from the CHY representation}},
  \href{https://doi.org/10.1016/j.physletb.2017.01.036}{\emph{Phys. Lett.}
  {\bfseries B767} (2017) 86--90},
  [\href{https://arxiv.org/abs/1607.06036}{{\ttfamily 1607.06036}}].

\bibitem{Fu:2017uzt}
C.-H. Fu, Y.-J. Du, R.~Huang and B.~Feng, \emph{{Expansion of
  Einstein-Yang-Mills Amplitude}},
  \href{https://doi.org/10.1007/JHEP09(2017)021}{\emph{JHEP} {\bfseries 09}
  (2017) 021}, [\href{https://arxiv.org/abs/1702.08158}{{\ttfamily
  1702.08158}}].

\bibitem{Teng:2017tbo}
F.~Teng and B.~Feng, \emph{{Expanding Einstein-Yang-Mills by Yang-Mills in CHY
  frame}}, \href{https://doi.org/10.1007/JHEP05(2017)075}{\emph{JHEP}
  {\bfseries 05} (2017) 075},
  [\href{https://arxiv.org/abs/1703.01269}{{\ttfamily 1703.01269}}].

\bibitem{Chiodaroli:2017ngp}
M.~Chiodaroli, M.~Gunaydin, H.~Johansson and R.~Roiban, \emph{{Explicit
  Formulae for Yang-Mills-Einstein Amplitudes from the Double Copy}},
  \href{https://doi.org/10.1007/JHEP07(2017)002}{\emph{JHEP} {\bfseries 07}
  (2017) 002}, [\href{https://arxiv.org/abs/1703.00421}{{\ttfamily
  1703.00421}}].

\bibitem{Johansson:2014zca}
H.~Johansson and A.~Ochirov, \emph{{Pure Gravities via Color-Kinematics Duality
  for Fundamental Matter}},
  \href{https://doi.org/10.1007/JHEP11(2015)046}{\emph{JHEP} {\bfseries 11}
  (2015) 046}, [\href{https://arxiv.org/abs/1407.4772}{{\ttfamily 1407.4772}}].

\bibitem{Johansson:2015oia}
H.~Johansson and A.~Ochirov, \emph{{Color-Kinematics Duality for QCD
  Amplitudes}}, \href{https://doi.org/10.1007/JHEP01(2016)170}{\emph{JHEP}
  {\bfseries 01} (2016) 170},
  [\href{https://arxiv.org/abs/1507.00332}{{\ttfamily 1507.00332}}].

\end{thebibliography}\endgroup
\bibliographystyle{/home/stefanw/latex-style/JHEP_new2}

\end{document}